\documentclass[pra,twocolumn,nofootinbib,showpacs,superscriptaddress]{revtex4-1}

\usepackage{graphicx}
\usepackage[dvips]{color}

\newcommand{\ket}[1]{\vert #1 \rangle} 
\newcommand{\braket}[2]{\langle #1 \vert #2 \rangle}
\newcommand{\ketbra}[2]{\vert #1 \rangle \! \langle #2 \vert}
\def\Tr{\hbox{tr}}

\begin{document}

\title{Statistical Estimation of the Quality of Quantum Tomography Protocols }
\author{Yu. I. Bogdanov}
\affiliation{Institute of Physics and Technology, Russian Academy of Sciences,
117218, Moscow, Russia}
\author{G. Brida}
\affiliation{INRIM, Strada delle Cacce 91 I-10135, Torino, Italy}
\author{I. D. Bukeev}
\affiliation{Institute of Physics and Technology, Russian Academy of Sciences,
117218, Moscow, Russia}
\author{M. Genovese}
\affiliation{INRIM, Strada delle Cacce 91 I-10135, Torino, Italy}
\author{K. S. Kravtsov}
\affiliation{Prokhorov General Physics Institute, Russian Academy of
Sciences, Moscow, 119991, Russia}
\author{S. P. Kulik}
\affiliation{Faculty of Physics, Moscow State University, 119992, Moscow,
Russia}
\author{E. V. Moreva}
\affiliation{Moscow National Research Nuclear University ``MEPHI'', 115409,
Moscow, Russia}
\author{A. A. Soloviev}
\affiliation{Faculty of Physics, Moscow State University, 119992, Moscow,
Russia}
\author{A. P. Shurupov}
\affiliation{INRIM, Strada delle Cacce 91 I-10135, Torino, Italy}

\date{\today}

\begin{abstract}
We present a complete methodology for testing the performances of quantum
tomography protocols. The theory is validated by several numerical examples and
by the comparison with experimental results achieved with various protocols for
whole families of polarization states of qubits and ququarts including pure,
mixed, entangled and separable.
\end{abstract}
 \maketitle

\section{Introduction}
Quantum information technologies rely on the use of quantum states in novel
data transmission and computing protocols \cite{Genovese, Nielsen,Scully}.
Control is achieved by statistical methods via quantum state reconstruction. At
present quantum state and process tomography serves as a principal instrument
for characterization of quantum states preparation and transformation quality
\cite{Ibort, Kwiat, Langford, Molina-Terriza, Hradil, Banaszek, Banaszektwo,
D'Ariano, Lvovsky, Zavatta,T, Allevi, Zambra, Bogdanov, Bogdanovtwo, krivitsky,
Chekhova, Moreva, DAriano, Rehacek, Ling, Lingtwo, Burgh, Asorey, Mikami,
Lanyon, Traina, Brida, teo, mat,mat2}.

In the last few years, a significant effort was made to improve tomographic
methods of quantum state measurements. In \cite{Nunn} a method of choosing the
best set of measurements from a finite list of possible ones is presented. The
basis of consideration for such problems is the asymptotic theory of
statistical estimations of the parameters of the density matrix. Difficulties
in solving optimization problems are caused by the large number of parameters
to be estimated. In general case, it requires extensive numerical calculations.
The number of possible measurements becomes extremely large if one wants to
cover a multidimensional parameter space of the measurement apparatus. Thus,
holding a finite number of samples in multidimensional parameter space, does
not guarantee that the truly optimal set of measurements is chosen. In other
words, even the method is universal and works for qubits as well as for higher
dimensional systems, it does not say which particular measurements have to be
performed. It just helps to choose the best measurements out of large
suboptimal set of possible measurements. Another significant problem is the
choice of an adequate parametrization for the quantum states (we will return to
this issue in Sec. \ref{sec:Discussion}).

The work \cite{Yan} provides a theoretical approach to optimal measurements of
qutrit systems, where {\em optimal} means the use of a minimal number of bases
for measurements. As expected, the use of mutually unbiased bases is {\em
optimal} in that sense, although its experimental realization, not discussed in
the paper, might be far from the optimal one.

Another recent work \cite{Toth} addresses the problem of density matrix
reconstruction for large multiqubit states, where a complete state
reconstruction is impossible due to the exponentially growing number of
undefined parameters with the increase of qubit number in the system. The
solution given is in finding only the permutationally invariant part of the
density operator, which gives a good approximation of the true quantum state in
many relevant cases.

In the present paper, we introduce a complete methodology for statistical
reconstruction of quantum states which is based on analysis of completeness,
adequacy and accuracy of quantum measurement protocols \cite{PRLlast,
JETPLlast, JETPLBogd, New}. Completeness of the quantum tomography protocol
provides the possibility of reconstruction of an arbitrary quantum states (both
pure and mixed) and it is assessed by means of singular value decomposition of
a special matrix built upon operators of measurement. Considered SVD allows us
to introduce the condition number $K$, which characterizes the quality of
quantum measurement protocol. Adequacy employs redundancy of a measurement
protocol compared to the minimum number of measurements required for
reconstruction. It is assessed by consistency of redundant statistical data and
mathematical model, based on quantum theory. Accuracy of statistical
reconstruction of quantum states is based on a universal statistical
distribution proposed in \cite{JETPLBogd}.

It is worth mentioning that in real experiments the accuracy of the state
reconstruction depends on two types of uncertainties: statistical and
instrumental ones. If the total number of measurement outcomes (sample size or
statistics) is large enough, the instrumental uncertainties dominate over
fundamental statistical fluctuations caused by the probabilistic nature of
quantum phenomena \cite{Bogdanov}. Practically, the required statistics,
allowing us to exclude statistical fluctuations, depends on the tomographic
protocol itself and the total accumulating time needed for taking data. From
this point of view it would be useful to point out simple and universal
algorithms for the estimation of the chosen protocol on the design stage before
doing experiments, as well as the sample size for desirable quality of the
state reconstruction.

The method considered in the present paper has the following features: (1) It
is well suited in the case of multi-qubit state tomography. (2) It accepts the
reconstruction for mixed states of arbitrary rank as well as for pure states.
(3) It allows to compare various quantum measurements protocols with each other
and, moreover, with respect to the fundamental fidelity level. (4) It also
motivates the experimenter to manage available resources in the best way as
well as to choose an optimal quantum measurement protocol. The paper is
organized as follows. Section~\ref{sec:Protocol} introduces the mathematical
apparatus for general quantum measurement protocol (for states with discrete
variables). In Sec.~\ref{sec:Distribution} we discuss a generalized statistical
distribution for fidelity by introducing specific random value which can be
called the loss of fidelity. Important operational criterion and corresponding
quantifier (so called conditional number) of the protocols' quality based on
completeness are suggested. Section~\ref{sec:Qubits} relates to analysis of
specific examples of tomographic protocols of qubits which are based on the
polyhedrons geometry. Section~\ref{sec:Results} collects results of numerical
and physical experiments for both pure and mixed states of single and pair of
qubits. Since it represents a paradigmatic example, here we restrict ourselves
to polarization degrees of freedom only; however, any other degrees of freedom
can be considered. In Sec.~\ref{sec:Discussion} we discuss in details the
features and advantages of the suggested approach in context of completeness,
adequacy and accuracy. Then we conclude with Sec.~\ref{sec:Conclude}.

\section{Quantum measurement protocol}\label{sec:Protocol}

An arbitrary $s$-dimensional quantum state is completely described by a state
vector in a $s$-dimensional Hilbert space when it is a pure state, or by a
density matrix $\rho$ for a mixed one. To measure the quantum state one needs
to perform a set of projective measurements on a set of identical states.

A quantum measurement protocol can be defined by a so-called instrumental
matrix $X$ that has $m$ rows and $s$ columns \cite{Bogdanov, Bogdanovtwo,
krivitsky}, where $s$ is the Hilbert space dimension and $m$ the number of
projections in such space. For every row, that is, for every projection, there
is a corresponding amplitude $M_j$,
\begin{equation}\label{eq:ampl}
M_{j}=X_{jl}c_{l}, \,\,\,\,\ j=1,2,\dots,m
\end{equation}
where we assume a summation by the joint index $l$, the $c_{l} (l =
1,2,\dots,s$) being the components of the state vector in the Hilbert space of
dimension $s$. The square of the absolute value of the amplitude defines the
intensity of a process, which is the number of events in 1 s
\begin{equation}\label{eq:prob}
\lambda_{j}=|M_{j}|^2.
\end{equation}
The number of registered events $k_{j}$ is a random variable Poissonially
distributed,  $t_{j}$ being the time of exposition of the selected row of the
protocol and $\lambda_j t_j$ the average value,
\begin{equation}\label{eq:poisson}
P(k_j) = \frac{{(\lambda_j t_j )^{k_j}}}{{k_j!}}\exp(-\lambda_j t_j).
\end{equation}
It is convenient to introduce special observables $\Lambda_j$ (so called
intensity operators), which are measured by the protocol during experiment,
\begin{equation}\label{eq:lambda}
\lambda_j = \Tr(\Lambda_j \rho)
\end{equation}
Here $\Lambda_j = X_j^\dag X_j$ is the intensity operator for the quantum
process $X_j$ (the row of the instrumental matrix $X$). In this case the
intensity operator for quantum process $\Lambda_j$ is a projector, so we have
\begin{equation}\label{eq:lambdasqr}
\Lambda_j^2 = \Lambda_j
\end{equation}
Formally, in the most general case, $\Lambda_j$ is an arbitrary positively
defined operator \cite{Holevo}. It can be presented as a mixture of projection
operators described above.
\begin{equation}\label{eq:lambdamix}
\Lambda _j =\sum\limits_k {f_k X_j^{(k) \dag } X_j^{(k)} }
\end{equation}
Here the index $k$ sums different components of the mixture that have weights
$f_k>0$. Such measurement can be conveniently presented as a reduction of the
set of projection measurements where only total statistics is available, while
statistical data for individual components are not available. The general
projection measurement is a particular case of Eq.~(\ref{eq:lambdamix}) where
$f_{1}=1,f_{2}=f_{3}=\cdots=0$. If the sum of the intensities multiplied by the
exposition time is proportional to a unit matrix, then we say that the protocol
is brought to a decomposition of unity \cite{Holevo},
\begin{equation}\label{eq:unitmatrix}
I = \sum\limits_{j = 1}^m {t_j \Lambda_j = I_0 E},
\end{equation}
where $I_0$ is the constant which defines overall intensity and $E$ is the
identity (or unit) matrix. A protocol for which the condition
(\ref{eq:unitmatrix}) holds in the general case can be brought to the so-called
non-orthogonal decomposition of the unity \cite{Holevo}. In this case the
protocol analysis is simplified. In mathematics such measurements are
considered as the most general extension of traditional von Neumann
measurements, which are based on the orthogonal decomposition. Even if it is
reasonable to require (\ref{eq:unitmatrix}) due to the total probability
preservation, it is worth mentioning that real experimental protocols often
cannot be brought to decomposition of unity. Indeed, in real experiments when
event registration scheme is used, the experimenter often adjusts his device to
distinguish only one projection of the quantum state (simply loosing data which
corresponds to the other states). Thereby real experiments (due to technical
requirements) usually do not provide registration of the whole statistical
ensemble and thus they are not restricted by total probability preservation
requirement. However, the method suggested in the present paper is applicable
to these cases as well.

The normalization condition for the protocol defines the total expected number
of events $n$ summarized by all rows:
\begin{equation}\label{eq:events}
\sum\limits_{j = 1}^m {\lambda_j t_j = n},
\end{equation}
where $t_j$ is the acquisition time. Condition (\ref{eq:events}) substitutes
the traditional normalization condition for the density matrices, $\Tr(\rho)=1$

In the following we consider the protocols of quantum measurements in terms of
two important notions--completeness and adequacy. For this purpose, we
introduce some opportune notation. First, when describing the whole sequence of
quantum protocol measurements each quantum process intensity matrix $\Lambda_j$
of dimension $s\times s$ is pulled into a single string of the length $s^2$ (to
put the second string to the right from the first etc). Then, we assign a
weight defined by an exposition time $t_j$ to each row in $B_j$ and we
construct a matrix $m \times s^2$ from these rows, calling it the measurement
matrix of the quantum protocol. We assume that $m \geq s^2$.

In the case of projection measurements defined by rows $X_j$ $(j=1,\dots,m)$ of
instrumental matrix $X$, the rows $B_j$ of the measurement matrix $B$ could be
calculated through the use of the tensor product of the row $X_j$ and the
complex conjugate row $X_j^\ast$.
\begin{equation}\label{eq:B}
B_j = t_j\cdot X_j^*\otimes X_j.
\end{equation}
In the following exposition times are assumed to be equal to 1. With this
matrix $B$, the protocol can be compactly written in the matrix form:
\begin{equation}\label{eq:K_e}
B\rho=T
\end{equation}
with $\rho$ being the density matrix, given in the form of a column (second
column lies below the first, etc.). The vector $T$ of length $m$ records the
total number of registered outcomes. The algorithm for solving Eq.
(\ref{eq:K_e}) is based on the so called singular value decomposition (SVD)
\cite{Kress}. SVD serves as a base for solving inverse problem by means of
pseudo-inverse or Moore-Penrose inverse \cite{Kress, Penrose}. In summary, the
matrix $B$ can be decomposed as:
\begin{equation}\label{eq:svd}
B=UVS^\dag,
\end{equation}
where $U$, $(m\times m)$, and $V$, $(s^{2}\times s^{2})$, are unitary matrices
and $S$, $(m\times s^{2})$, is a diagonal, non-negative matrix, whose diagonal
elements are ``singular values''. Then (\ref{eq:K_e}) transforms to a simple
diagonal form:
\begin{equation}\label{eq:S}
Sf=Q
\end{equation}
with a new variable $f$, unitary related to $\rho$ via $f=V^\dag\rho$, and a
new column $Q$, unitary related to the vector $T$ by the equation $Q=U^\dag T$.
This system is easy to solve because $S$ is a diagonal matrix. Its analysis
allows classifying measurements from the viewpoint of adequacy and completeness
\cite{New}.

Let $m>s^{2}$, that is, the number of measurements is greater than the number
of elements in the density matrix. The rank of the model $q$ denotes the number
of non-zero singular values of the matrix $B$. By defining $q$ we formulate two
important conditions of any tomography protocol, namely its completeness and
adequacy \cite{New}. It is obvious that $q\leq s^2$. The last $m-q$ rows in the
matrix $S$ are equal to zero. Then it follows that for the system to be
adequate, it is necessary that the last $m-q$ values in the characteristic
column $Q$ are also equal to zero. We will refer to this condition as a
measurements adequacy condition. If it does not hold the model is inadequate,
that is, statistical data do not correspond to any quantum mechanical density
matrix. It may mean, for example, that either the experiment is realized
incorrectly or measurement matrix is wrong. Adequacy means that the statistical
data directly correspond to the physical density matrix (which has to be
normalized, Hermitian and positive). However it is worth noticing that,
generally, for mixed state it can be tested only if the protocol consists of
redundant measurements (i.e. if $m>s^2$).

Suppose that the model is adequate.  The protocol is supposed to be
informational complete if the number of tomographically complementary
projection measurements is equal to the number of parameters to be estimated;
mathematically completeness means $q=s^2$. If all singular values are knowingly
nonzero, that is, $q=s^2$, then unconditional completeness holds and a solution
exists and is unique. Measurement protocol completely returns any quantum state
(pure and mixed) that could be defined in the considered Hilbert space.

In this case we can determine the factor column dividing the elements of
characteristic column by the corresponding singular values
\begin{equation}\label{eq:f}
f_{j}=Q_{j}/S_{j}\,\,\,\,\ j=1,2,\dots,s^2,
\end{equation}
As a result we obtain the desired density matrix by a unitary transformation:
\begin{equation}\label{eq:ro}
\rho=Vf.
\end{equation}
Due to the unitarity of the matrix $V$, the factor column $f$ determines the
degree of purity of the quantum state:
\begin{equation}\label{eq:tr_ro}
\Tr(\rho^2) = \sum\limits_{j=1}^m {\left| {f_j} \right|^2}.
\end{equation}
Finally, suppose $q<s^{2}$, that is, some singular values are equal to zero. In
this case for nonzero values we have:
\begin{equation}\label{eq:f_1}
f_{j}=Q_{j}/S_{j}, \,\,\,\ j=1,2,\dots,q.
\end{equation}
Let us call these factors $f_j,\,j=1,2,\dots,q$ defined factors.

At the same time, for nonzero values, we have equations corresponding to
uncertainty ``zero divided by zero''
\begin{equation}\label{eq:zero_factor}
0f_{j}=0,\,\,\,\,\ j=q+1,\dots,s^2.
\end{equation}

We shall call these factors $f_{j}\,\,\,\, j=q+1,\dots,s^2$ as undefined
factors. As solutions of the last equations arbitrary complex numbers could be
used. The considered situation corresponds to the incompleteness of
measurements and this system of equations has an infinite number of solutions.
However, not all of them correspond to real physical density matrices. The
physical solutions only correspond to that having a Hermitian nonnegative
definite density matrix. Formally all these solutions could be obtained by
scanning all possible values of undefined factors. It is evident that such a
procedure could be carried out only when the dimension of undefined factors'
space is relatively small.

Let us call regularized (normal) a solution corresponding to the special choice
of all undefined factors: $f_j=0$ for $j=q+1,\dots,s^{2}$. Due to the unitarity
of relation between the density matrix and the factor column, a regularized
solution corresponds to the minimum purity level of the reconstructed state. In
this case we have:
\begin{equation}\label{eq:eq}
\Tr(\rho^2) \geq \sum\limits_{j=1}^m {\left| {f_j} \right|^2}.
\end{equation}
One could see that in the case of incomplete protocols any additional
measurements could either cause an increase of the purity or leave it
unchanged. If the regularized solution already describes a pure state, then new
measurements will not influence the reconstructed state. In other words in this
case we can obtain complete information about the considered state despite of
incomplete measurement protocol. We will call such protocols conditionally
complete, that is, there is completeness under condition that only specially
chosen states are considered; for instance, this is the case of experiments
like ``which way'' \cite{Valiev}.

Then we assume that there is unconditional completeness ($q=s^2$). Equations
(\ref{eq:f}) and (\ref{eq:ro}) could be used to approximate the reconstruction
of the density matrix if we substitute experimental events (frequencies) into
the right side of Eq. (\ref{eq:K_e}). However due to statistical fluctuations
of experimental data the reconstructed matrix will not always be positive
definite (components with small weights could be reconstructed as negative
numbers in this case). Despite this disadvantage, the method provides a good
zero approximation for widely used maximum likelihood method (ML). In this
case, the components with negative weights simply assumed to be zero, then the
density matrix is multiplied by a factor that ensures the correct
normalization. Note that ML method itself is free from the considered
disadvantage because positive definiteness lies in the nature of the method. At
the same time the zero approximation, obtained from pseudo inversion method,
significantly accelerates the search of ML solution.

\section{Universal statistical distribution for fidelity losses and the maximum
possible fidelity of quantum states reconstruction}\label{sec:Distribution}

The accuracy of quantum tomography can be defined by a parameter called
fidelity \cite{Nielsen, Uhlmann}
\begin{equation}\label{eq:fidelity}
F = \left[ \Tr\sqrt{ \sqrt{\rho^{(0)}} \rho \sqrt{\rho^{(0)}} } \right]^2,
\end{equation}
where $\rho^{(0)}$ is theoretical density matrix and $\rho$ is reconstructed
density matrix. The fidelity shows how close the reconstructed state is to the
ideal theoretical state: the reconstruction is precise if the fidelity is equal
to one.

This equation looks quite complex, but it becomes simple if we apply the
Uhlmann theorem \cite{Uhlmann}. According to this theorem, the Fidelity is
simply the maximum possible squared absolute value of the inner product:
\begin{equation}\label{eq:fidelity_pure}
F=\left| \braket{c_0}{c} \right|^2
\end{equation}
where $c_0$ and $c$ are theoretical and reconstructed purified state vectors.

We explicitly use the Uhlmann theorem in our algorithm of statistical
reconstruction of quantum states, based on maximum likelihood method. This fact
is very important even if the state is not pure, we have to purify it by moving
into a space of higher dimension \cite{JETPLBogd}.

It is well known that purified state vectors are defined ambiguously. However,
this ambiguity does not preclude from reconstructing a quantum state, which is
a very useful feature of the suggested algorithm. It is devised in the way that
different purified state vectors produce the same density matrix and therefore
the same fidelity during the reconstruction. This is a key principle for the
proposed procedure and thus reconstruction can be obtained by means of
purification. Purification greatly facilitates the search of a solution,
especially when we need to estimate a great number of parameters (hundreds or
even thousands).

The fidelity level (\ref{eq:fidelity_pure}) has a simple probabilistic
interpretation. If we choose a known reconstructed vector and its orthogonal
complement as a measurement basis for an unknown state $c_0$ then $F$ returns
the probability that this unknown state coincides with the reconstructed one.

It is equally important that due to the usage of purification procedure we
succeed in formulating a generalized statistical distribution for fidelity
\cite{JETPLBogd}. Purification procedure will be described below in
Sec.~\ref{sec:Discussion} [formula (\ref{eq:LL}) provides with a transition
from the density matrix to a purified state vector]. The value $1-F$ can be
called the loss of fidelity. It is a random value and its asymptotical
distribution can be presented in the following form:
\begin{equation}\label{eq:average_losses}
1-F =\sum_{j=1}^{j_{\text{max}}}d_j\xi^2_j
\end{equation}
where $d_{j}\geq 0$ are non-negative coefficients, $\xi_{j} \sim N(0,1)$
$j=1,\dots,j_{\text{max}}$ are independent normally distributed random values
with zero mean and variance equal to one, $j_{\text{max}}=(2s-r)r-1$ is the
number of degrees of freedom of a quantum state and corresponding distribution,
$s$ is the Hilbert space dimension, and $r$ is the rank of mixed state, which
is the number of non-zero eigenvalues of the density matrix.  In particular
$j_{\text{max}}=2s-2$ for pure states and $j_{\text{max}}=s^2-1$ for mixed
states of full rank ($r=s$).

This distribution is a natural generalization of the $\chi^2$ distribution.
Ordinary $\chi^2$ distribution corresponds to the particular case when
$d_{1}=d_{2}=\cdots=d_{j_{\text{max}}}=1$ (all components of vector $d$ are
equal to one). In the asymptotic limit considered by us, the parameters $d_{j}$
are inversely proportional to the sample size $n$, that is,
$d_{j}\sim\frac{1}{n}$. This dependence allows for an easy recalculation when
we use different sample sizes.

The method of calculating the parameter vector $d$ is based on Fischer's
information matrix. In this case, the eigenvectors of information matrix define
directions of principal fluctuations of the purified state vector whereas the
respective variances of the principal fluctuations are inversely proportional
to the eigenvalues of the information matrix. The method is described in
\cite{JETPLBogd} in details.

From Eq. (\ref{eq:average_losses}) one gets the average fidelity loss
\begin{equation}\label{eq:aver}
\left\langle {1-F} \right\rangle  = \sum\limits_{j=1}^{j_{\text{max}}} {d_j}.
\end{equation}
It is also easy to show that the variance for the fidelity loss is
\begin{equation}\label{eq:sigma}
\sigma^2 = 2\sum\limits_{j=1}^{j_{\text{max}}} {d_j^2}.
\end{equation}
Moments of higher order for this distribution can be calculated analytically.
For example the momentum of third order is called skewness and describes, for a
random variable $x$, the asymmetry
\begin{equation}\label{eq:beta1_2}
\beta_1 = \frac{ M\{[x-M(x)]^3\} }{ \sigma^3 }
\end{equation}
$M$ denoting the mathematical expectation. The fourth-order moment
is called excess kurtosis,
\begin{equation}\label{eq:beta2_2}
\beta_2 = \frac{ M\{[x-M(x)]^4\} }{\sigma^4} - 3
\end{equation}
The equations for $\beta_1 ,\beta_2$ are:
\begin{equation}\label{eq:beta1}
\beta_1 = \frac{ 8\sum\limits_{j=1}^{j_{\text{max}}} {d_j^3} }{\sigma^3}
\end{equation}
\begin{equation}\label{eq:beta2}
\beta_2 = \frac{ 48\sum\limits_{j=1}^{j_{\text{max}}} {d_j^4} }{\sigma^4}
\end{equation}

Let us consider a special case of the quantum state, which is defined by a
uniform density matrix. Moreover, this matrix is proportional to the identity
matrix. Such state can be represented as a ``white noise'' for which all
weights of principal components are equal. Let us assume that the protocol can
be brought to projection measurements, which form non-orthogonal decomposition
of unity in accordance with (\ref{eq:lambdasqr}) and (\ref{eq:unitmatrix}). In
this case there is a simple relation between the vector $d$, of size $s^{2}-1$
determining the distribution of losses and the vector of the singular values
for the measurement matrix $B$ (the element of the greatest value should be
eliminated from this vector of size $s^{2}$). Denoting the reduced vector of
the size $s^{2}-1$ as $b$, one gets the relation between these vectors:
\begin{equation}\label{eq:d}
d_j = \frac{C} {{nb_j^2 }},\,\,\ j = 1,2,\dots,s^2-1,
\end{equation}
where the constant $C$ is given by
\begin{equation}\label{eq:C}
C = \frac{\sum\limits_j {b_j^2} }{4(s - 1)}.
\end{equation}
For multiqubit protocols considered in the next section, which are based on
polyhedrons, we have:
\begin{equation}\label{eq:d_j}
d_j = \frac{m^l}{4snb_j^2},
\end{equation}
where $m$ is the polyhedron's faces number and $l$ is a number of qubits in the
register. It is obvious from (\ref{eq:d_j}) that
\begin{equation}\label{eq:ratio_d}
\frac{d_{\text{max}}}{d_{\text{min}}} = \left( \frac{b_{\text{max}}}
{b_{\text{min}}} \right)^2.
\end{equation}
In addition, it can be shown that for such protocols the following equation holds:
\begin{equation}\label{eq:ratio_1}
\frac{b_{\text{max}}} {b_{\text{min}}} = \left( \sqrt3 \right)^{l-1}.
\end{equation}
In this case the condition number $K$ of the matrix $B$ is
\begin{equation}\label{eq:cond_B}
K=\text{cond}(B) = \left( \sqrt3 \right)^l.
\end{equation}
Recall that the condition number of a matrix is the ratio of the maximum
singular value to the minimum one. Note also that in the definition
(\ref{eq:cond_B}) all singular values are taken into account while in the
(\ref{eq:ratio_1}) the one is ignored due to normalization.

Let us introduce a value of fidelity loss, which is independent on the sample
size.
\begin{equation}\label{eq:L}
L = n\left\langle {1-F} \right\rangle = n\sum\limits_{j=1}^{j_{\text{max}}}
{d_j}.
\end{equation}
This quantity serves as the main figure of merit of the precision in the
examples analyzed below: the lower the value of the loss function (\ref{eq:L}),
the higher is the precision of the protocol. As we can see from the Eq.
(\ref{eq:L}), this value is determined by vector $d$ that defines the general
fidelity distribution. Therefore, the general fidelity distribution serves as a
tool for completely solving the problem of precision for quantum tomography.

As an important example, let us consider the protocol defined by projection
measurements, which form non-orthogonal decomposition of the unity. It can be
shown that the following condition holds in this case
\begin{equation}\label{eq:frac}
\frac{1}{4n}\sum\limits_{j=1}^\nu {\frac{1}{d_j}} = s-1
\end{equation}
Here $\nu$ is the number of parameters to define a state:
\begin{equation}\label{eq:degrees}
\nu = j_{\text{max}} = (2s-r)r-1
\end{equation}
Optimization of the problem leads to search of the minimum level of fidelity
losses (\ref{eq:average_losses}) such that (\ref{eq:frac}) holds. Obviously the
problem is solved when $d_1=d_2=\cdots=d_\nu$, when mean losses approach their
minimum level:
\begin{equation}\label{eq:min_fidelity}
\left\langle {1-F} \right\rangle_{\text{min}} = \frac{\nu^2} {4n(s-1)}
\end{equation}
Note that the requirement $d_{1}=d_{2}=\cdots=d_{\nu}$ not only defines a
minimum level of mean losses (\ref{eq:aver}), but also a minimum of other
moments of losses [variance (\ref{eq:sigma}), skewness (\ref{eq:beta1}) and
excess kurtosis (\ref{eq:beta2})].

It appears from (\ref{eq:min_fidelity}) that the minimum possible loss is given
by the following equation:
\begin{equation}
L_{\text{min}}^{\text{opt}} = \frac{\nu^2}{4(s-1)}.
\end{equation}
For pure states $\nu=2s-2$, the possible loss is given by:
\begin{equation}
L_{\text{min}}^{\text{opt}} = s-1.
\end{equation}
For mixed states of full rank $\nu=s^2-1$ the loss is given by the following
equation:
\begin{equation}
L_{\text{min}}^{\text{opt}} = \frac{(s+1)^2 (s-1)}{4}.
\end{equation}
Any protocol for any quantum state cannot have losses lower that
those defined by this equation, if the protocol can be brought to
projection measurements, which form non-orthogonal decomposition of
the unity. Note that if the protocol cannot be brought to
decomposition of unity then the losses can be lower than defined by
this equation. However, this is true only for certain states and an
improvement in reconstruction precision for some states is
completely compensated by a significant deterioration in
reconstruction precision for other states.

\section{Protocols based on the polyhedrons geometry}\label{sec:Qubits}

In this section we present analysis of specific examples of tomographic
protocols which are based on the polyhedrons geometry. The multiqubit protocols
described in this section are formed by projective quantum measurements on
states that are tensor products of single-qubit states. For example, if a
single-qubit measurement protocol is formed by a projection onto a polyhedron
with $m$ faces inscribed in the Bloch sphere, then it has $m$ rows (see below).
Therefore, the corresponding $l$-qubit protocol possess $m^l$ rows.

Intuitively, projection of the state under consideration onto symmetric solids
inscribed in the Bloch sphere leads to better accuracy of the reconstruction.
Shown below is a study based on the calculation of the loss function
(\ref{eq:L}), which significantly extends the results obtained in \cite{Burgh,
Rehacek}. In the term of single-qubit protocols it is worth highlighting
regular polyhedrons and polyhedrons with lesser but still rather high level of
symmetry.

Regular polyhedrons or Platonic solids are used for the most symmetrical and
uniform distribution of quantum states on the Bloch sphere. Projections are
defined by directions from the center of Bloch sphere to centers of polyhedron
faces. Therefore, the number of polyhedron's faces defines the number of
protocol's rows and is equal to 4 for tetrahedron, 6 for cube, 8 for
octahedron, 12 for dodecahedron and 20 for icosahedron.

These five bodies form the complete set of regular polyhedrons. The search of
quantum measurements protocols with high symmetry on Bloch sphere and number of
rows greater than twenty requires to consider non-regular polyhedrons, which
have high symmetry. As examples of such polyhedrons we have chosen fullerene
(truncated icosahedron) that defines quantum measurement protocol with 32 rows
(equal to the number of fullerene's faces) and also a dual~\footnote{A pair of
polyhedra are called dual, if the vertices of one correspond to the faces of
the other.} to fullerene polyhedron (pentakis dodecahedron) which defines
quantum measurement protocol with 60 rows (that is the number of its faces and
also the number of vertices of fullerene).

It is noteworthy that all protocols considered here can be brought to
decomposition of the unity [36].

A comparison of the maximal possible fidelity with the fidelity of the
protocols considered here shows that as the number of polyhedrons' faces
increases fidelity rapidly converges to the theoretical limit (in addition,
rapidly increases a uniformity of fidelity distribution on the Bloch sphere).
We should mention that the accuracy of the suggested protocols is much higher
in comparison with protocols exploiting not so highly symmetrical states.

As an example we calculated numerically a set of pictures that demonstrates the
Bloch sphere scanning by means of various measurement protocols for the single
qubit pure states. The corresponding colors indicate the value of the Fidelity
loss function.

\begin{figure}[t]
\includegraphics[width=1\columnwidth]{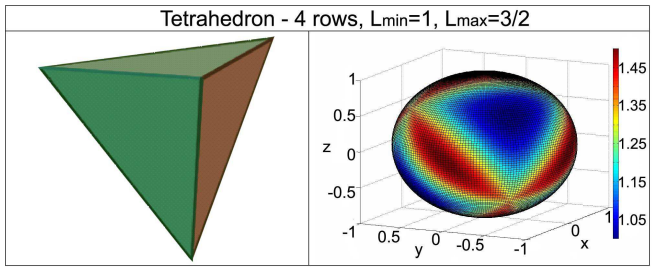}
\caption{(Color online) Tetrahedron and distributions fidelity loss over the
Poincar\'e-Bloch sphere for the protocol based on tetrahedron geometry. Color
bar shows level of average fidelity loss $L$.} \label{f:tetrahedron}
\end{figure}

Figure~\ref{f:tetrahedron} defines the value of the loss function for the
protocol based on tetrahedron. The minimal losses are equal to 1 as well as for
other figures. The maximum losses for tetrahedron are equal to 3/2.

\begin{figure}[b]
\includegraphics[width=1\columnwidth]{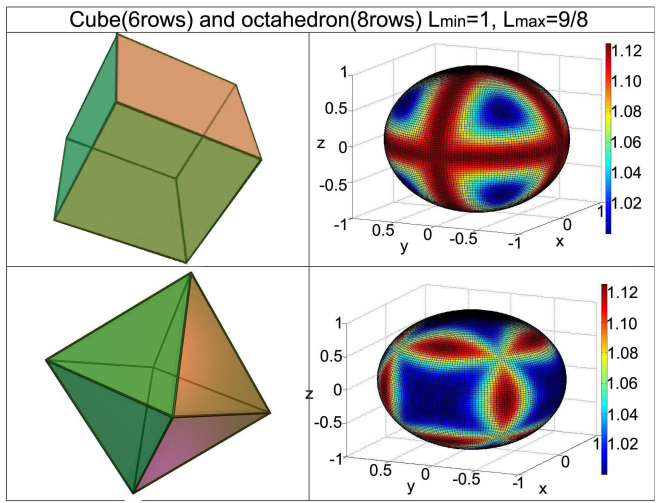}
\caption{(Color online) Shapes of solids and distributions of average fidelity
loss over the Poincar\'e-Bloch sphere for the protocol based on cube and
octahedron geometry. Color bar shows level of average fidelity loss $L$.}
\label{f:cube}
\end{figure}

Figure~\ref{f:cube} presents a cube and an octahedron. These polyhedrons are
dual to each other. The maximum losses are equal to 9/8 in both cases.

\begin{figure}[t]
\includegraphics[width=1\columnwidth]{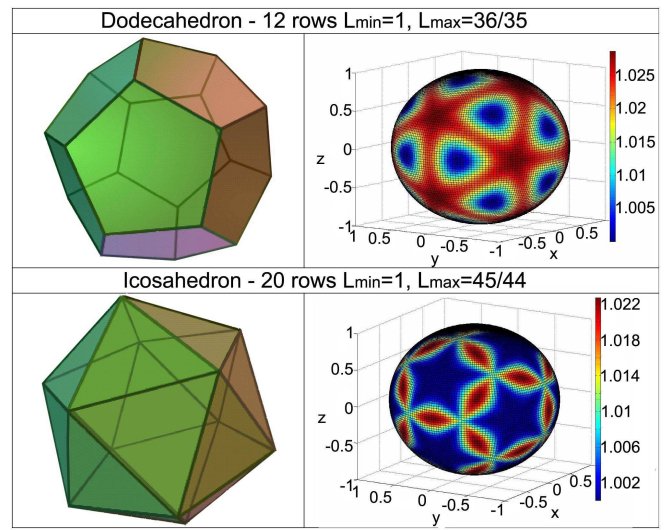}
\caption{(Color online) Shapes of solids and distributions of average fidelity
loss over the Poincar\'e-Bloch sphere for the protocol based on dodecahedron
and icosahedron. Color bar shows level of average fidelity loss
$L$.}\label{f:icosahedron}
\end{figure}

\begin{figure}[b]
\includegraphics[width=1\columnwidth]{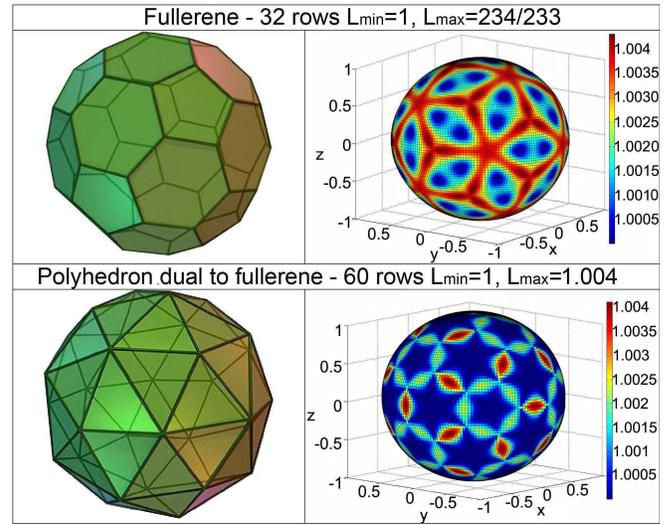}
\caption{(Color online) Shapes of solids and distributions of average fidelity
loss over the Poincar\'e-Bloch sphere for the protocol based on fullerene and
its dual polyhedron. Color bar shows level of average fidelity loss
$L$.}\label{f:fullerene}
\end{figure}

Figures \ref{f:icosahedron} and \ref{f:fullerene} present a dodecahedron and an
icosahedron, as well as fullerene and a polyhedron that is dual to the latter:
one can clearly see that when the number of projections grows the maximum
losses converge to the minimum possible losses. In the limit of infinite number
of points on the Bloch sphere we get an optimal protocol for which the
precision of reconstruction does not depend on the reconstructed state at all.
The price for that would be an infinite growth of the performed measurements,
so in a real experiment one should sacrifice the fidelity losses (accuracy) to
use a realistic (limited) number of measurements. However, this point is quite
common for any quantum tomography protocols exploiting redundant (with respect
to dimension of the reconstructed state) number of measurements.

For the sake of completeness, fullerene and its dual protocols are presented on
the last picture, while Table~\ref{t:losses} presents the results of numerical
experiments for pure quantum states with the number of qubits from 1 to 3. It
is worth noting that the algorithm of numerical optimization does not garantee
the finding of the global optima. Using the numerical procedure we have only
found hypothetical maximum loss values.

The precision of each protocol can be characterized by the following bounds
$L_{\text{min}} \le L \le L_{\text{max}}$. This inequality defines a rather
narrow range, where the precision of quantum state reconstruction is localized
definitely.

Numerical calculations demonstrate that the minimum possible losses
$L_{\text{min}}$ are defined by the theoretically derived optimal limit
$L_{\text{min}} = L_{\text{min}}^{\text{opt}}= s-1$.

By looking to the upper limit $L_{\text{max}}$, shown in the
Table~\ref{t:losses}, as the result of numerical experiments, one can evince
that for single-qubit protocols as the number of projections grows the maximum
losses converge to the minimum possible losses. In the limit of an infinite
number of points on the Bloch sphere one gets an optimal protocol for which the
precision of the reconstruction does not depend on the reconstructed state.

However, this is not true for multiqubit protocols. For two-qubit protocols,
the maximum possible losses approach the level $L_{\text{max}} \approx 3.38$
while the minimum possible level is equal to $L_{\text{min}}^{\text{opt}}=3$.
Similarly for three-qubit states the values are $L_{\text{max}} \approx 7.7$
and $L_{\text{min}}^{\text{opt}}=7$. These results are due to the fact that
these protocols are based on projections only onto non-entangled states.

\begin{table*}[t]
\caption{Results of numerical experiments that define maximum precision losses
$L_{\text{max}}$ for protocols based on the pohyhedron geometry.}
\begin{tabular}{lccc}\hline\hline
& 1 qubit & 2 qubits & 3 qubits\\
& ($s=2$, $L_{\text{min}}  = 1$) &  ($s=4$, $L_{\text{min}}  = 3$) & ($s=8$, $L_{\text{min}}  = 7$) \\
\hline
Tetrahedron ($m=4$) & 3/2=1.5 & 4.442971458 & $\approx 10.4$\\
Cube ($m=6$) & 9/8 = 1.125 & $\approx 3.5839$ & $\approx 8.2$\\
Octahedron ($m=8$) & 9/8=1.125 & 3.4708(3) & $\approx 7.9$\\
Dodecahedron ($m=12$) & 36/35 & $\approx3.42$ & $\approx 7.8$\\
Icosahedron ($m = 20$) & 45/44 & $\approx 3.39$ & $\approx 7.8$\\
Fullerene ($m=32$) & $\approx 234/233$ & $\approx 3.38$ & $\approx 7.7$\\
Polyhedron dual to fullerene ($m=60$) & 1.0041037488 & $\approx 3.38$ & $\approx 7.7$\\
\hline\hline
\end{tabular}
\label{t:losses}
\end{table*}

Here we should stress again that from the theoretical point of view in the
multiqubit case the discussed protocols are not the best possible ones because
they do not involve projections onto entangled states. In that case the
precision will be somewhat smaller than the minimum possible limit.
($L_{\text{max}} \approx 3.38$ compared to $L_{\text{min}}^{\text{opt}}=3$ for
two-qubit states and $L_{\text{max}} \approx 7.7$ compared to
$L_{\text{min}}^{\text{opt}}=7$ for three-qubit states).

It is also worth noting that, though the protocols based on polyhedrons with
small number of faces (tetrahedron, cube) are somewhat less precise, they are
much easier in practical implementation.

When considering tomography of mixed states it is worth noting that there is no
finite upper limit for precision losses (losses can be infinitely large $L\to
\infty$). Such large losses are inherent to mixed states, which are close to
pure ones. In fact the number of real parameters that define a mixed state of
full rank in Hilbert space of dimension $s$ is equal to $s^2-1$, which is
significantly greater for large $s$  than for a pure state that takes only
$2s-2$ real parameters. In case of a mixed state that has one predominant
component, the smaller weight components almost do not affect the statistical
data and do not increase the amount of Fisher information for reconstructing
the greatly larger number of parameters. Theoretical analysis, numerical and
real physical experiments completely prove these statements.

The lower limit for precision losses can be applied to mixed states as well. In
this case optimal minimum losses are realized for ``white noise states''
(uniform density matrix), when all components have equal weights. Then, there
is a simple relation between the vector $d$ of dimension $s^2-1$,  that defines
distribution of precision loss, and the vector of singular values of the
measurement matrix $B$. Corresponding estimate for $L_{\text{min}}$ for the
protocols considered here is given by the following equation:
\begin{equation}\label{eq:Lmin1}
L_{\text{min}} = \left(n\sum_j d_j\right)_{\text{min}} =
\sum_j\frac{m^l}{4sb_j^2} = \frac{10^l - 1}{4}.
\end{equation}
This value depends on the number of qubits, but does not depend on the type of
polyhedron. It defines the minimum possible losses for the considered protocols
that do not use projections on entangled states.

Recall that in the general case for any protocols including those ones that
involve projections onto entangled states minimum (optimal) losses are
described by:
\begin{equation}\label{eq:Lmin2}
L_{\text{min}}^{\text{opt}} = \frac{\nu^2}{4(s-1)} =
\frac{(2^l+1)^2(2^l-1)}{4}.
\end{equation}

By comparing the Eqs. (\ref{eq:Lmin1}) and (\ref{eq:Lmin2}) one can see that
these protocols provide minimum possible (optimal) losses for reconstruction of
mixed states of full rank only for single-qubit states. Therefore, for
multi-qubit cases protocols that provide minimum possible losses during quantum
states reconstruction should necessarily include projections onto entangled
states.

\begin{figure}
\includegraphics[width=1\columnwidth]{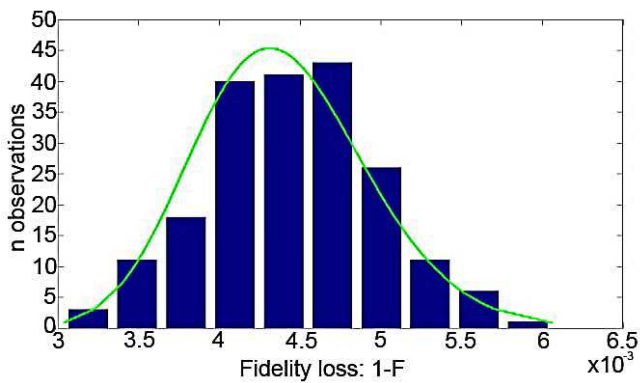}
\caption{(Color online) The universal statistical fidelity loss distribution
for 200 numerical experiments (histogram) compared with the theoretical
expectation (solid curve). Sample size of each experiment is 1 million.}
\label{f:modulation}
\end{figure}

Let us consider few examples demonstrating the features of the developed
approach. Figure~\ref{f:modulation} presents the results of numerical
experiments testing the universal statistical distribution for fidelity; 200
experiments were conducted with sample size 1 million each.

The measurement protocol is based on tetrahedron. We considered a
four-qubit state that represents a mixture of GHZ state and uniform
density matrix (white noise):

\begin{equation}
\rho = f\frac{E}{16} + (1-f)\ketbra{\text{GHZ}}{\text{GHZ}},
\end{equation}
where $E$ is the unit matrix of size $16\times16$, $\ket{\text{GHZ}}$ is the
state of Greenberger-Horne-Zeilinger: $\ket{\text{GHZ}} = \frac1{\sqrt 2}
(\ket{0000}+\ket{1111})$, and $f$ is the weight of the uniform density matrix
(white noise).  In our case $f=0.5$ (50\%). It is a multiparametric
distribution, being the size of the vector of parameters 255. These data
(Fig.~\ref{f:modulation}) demonstrate a good agreement between results of a
numerical experiment and the theory with high critical significance level
(0.65) for $\chi^2$ criterion.

Then we consider the dependence of reconstruction precision on the weight of
the ``white noise'' component. The value of fidelity can belong to a wide
interval, thus it is convenient to use a new variable $z=-\log (1-F)$. Here and
below $\log$ denotes the common logarithm. The new variable $z$ defines the
number of nines in numerical representation of fidelity, for example, $z=3$
means that $F=0.999$.

\begin{figure}
\includegraphics[width=1\columnwidth]{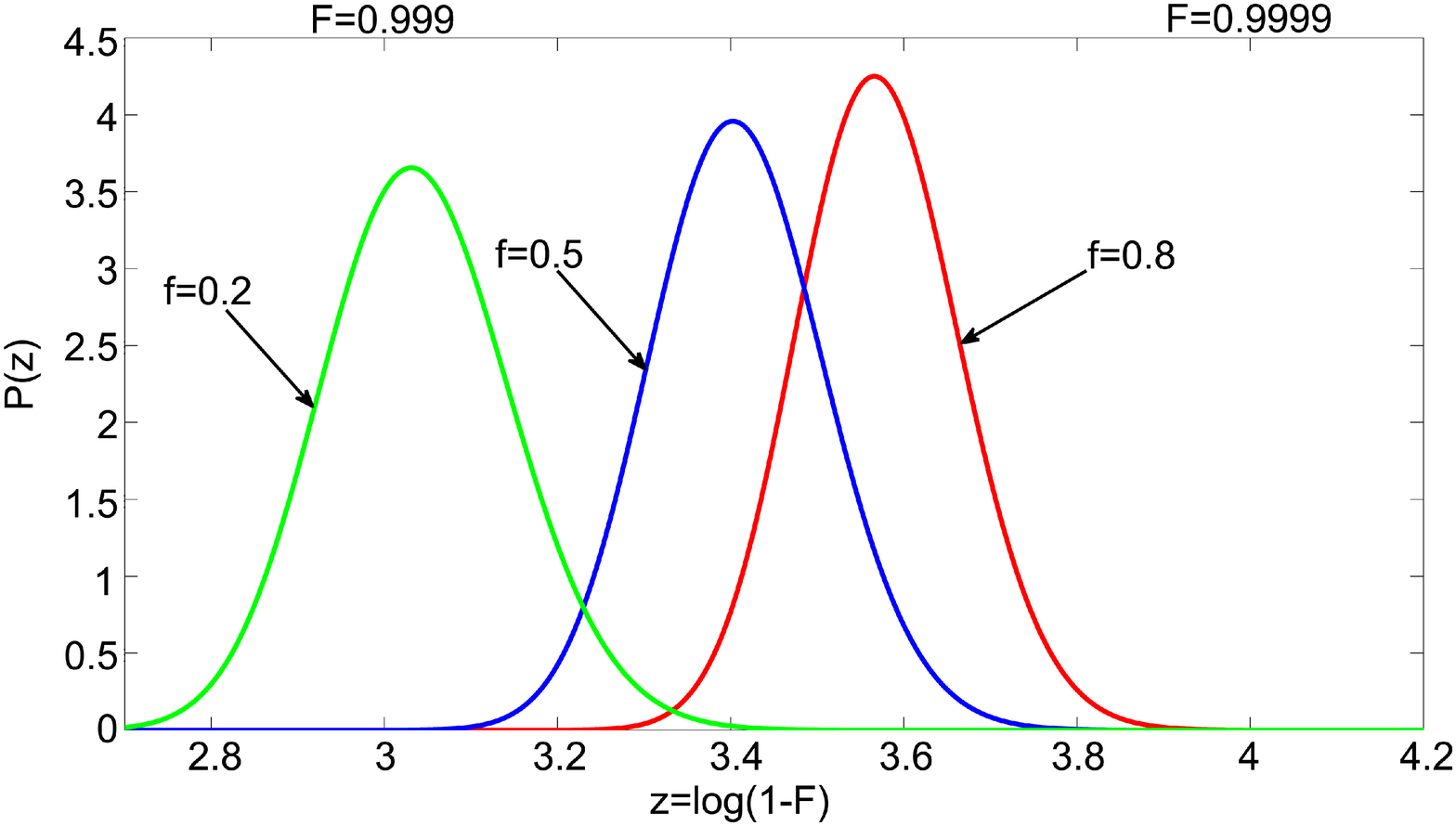}
\caption{(Color online) Density distribution of the scaled fidelity $z$ for
three qubit state with different value of mixture GHZ and ``white
noise''states. The measurement protocol is based on dodecahedron. Sample size
$n$ is equal to one million.} \label{f:fid_distr}
\end{figure}

As an another example, Fig.~\ref{f:fid_distr} shows distributions calculated by
using the new variable for a three-qubit state, which is again a mixture of GHZ
state and ``white noise'' (uniform density matrix). Calculations were performed
with a protocol based on dodecahedron. The sample size $n$ is equal to one
million as well. It is evident that the higher the white noise weight the
higher the precision of reconstruction. It is not difficult to explain this
fact in the context of previous discussion. Namely  the ``white noise'' is the
best one for mixed state reconstruction.

\begin{figure}
\includegraphics[width=1\columnwidth]{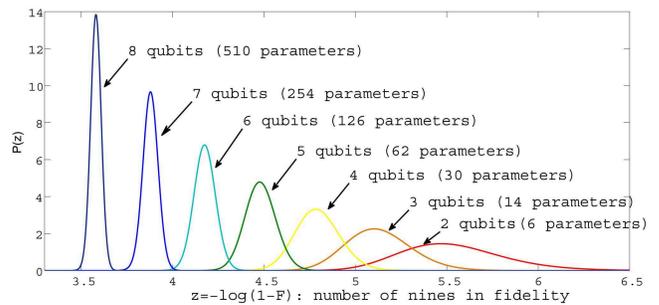}
\caption{(Color online) Density distribution of the scaled fidelity $z$ for
Bell and GHZ states. The measurement protocol is based on tetrahedron. Sample
size $n$ is equal to one million.} \label{f:Bell_GHZ}
\end{figure}

And finally, Fig.~\ref{f:Bell_GHZ} presents the behavior of the reconstruction
precision for Bell and GHZ states when the number of qubits grows from two to
eight and the sample size is one million. The measurement protocol is based on
tetrahedron. It is evident that the precision of reconstruction falls as
rapidly as the width of the distribution with increasing the number of qubits.

\section{Experiment and analysis}\label{sec:Results}

For testing the theoretic approach, described in the previous sections, we have
prepared a whole family of polarization states for qubits and ququarts. In
particular we considered both pure and mixed qubit states and pure, mixed,
entangled and separable states of ququarts.

\begin{figure}[ht]
\includegraphics[width=1\columnwidth]{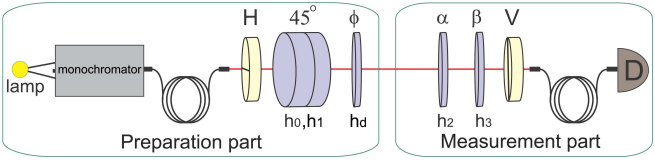}
\caption{(Color online) Experimental setup for different tomographic
reconstructions of qubits with variable degree of mixture. Here H, V are
Glan-Thompson prisms, $h_0$ is a thick quartz plate, $h_1$ is a thick tilt
plate, $h_d$ is a driving phase plate, $h_{2,3}$ are tomographic phase plates,
and D is a detector.} \label{f:setup_qubits}
\end{figure}

\textit{Qubits.} The experimental setup for generation and measurement of qubit
states is shown in Fig.~\ref{f:setup_qubits}. Setting up the experiment we
pursuit the following goals: (1) The setup should allow  performing the
transition between pure and mixed polarization states of qubits. (2) Three
different measurement protocols (R4, K4, B36) should be realized. (3) The setup
should allow performing measurements with different sample size (statistics).

To satisfy the first request we used a wide-band light source (incandescence
lamp) passing through a monochromator and thick birefringent plates. This
provides a variable spectral range around the chosen central wavelength
1.55~${\mu}$m and a controllable phase delay between basic polarization modes.
Therefore, by changing the spectral range of the light (within 1--23~nm) we
were able to vary the purity of the output polarization state. A parallel light
beam was formed by a $SMF28$ single-mode fiber with $F240FC$-1550
micro-objectives placed at its input and output. A Glan-Thompson prism was used
for preparation of the horizontally polarized state $\ket{H}$ serving as
initial state for the following transformations. As a result, the original pure
state was transformed into a mixed state with a degree of purity, depending on
the spectral width of the detected radiation. A smooth transition from pure
states to total mixtures was achieved with increasing spectral width of
radiation. Since the ``measurement part'' of the setup had a finite spectral
band, the polarization states at different spectral components within this band
were integrated, which corresponded to the registration of mixed-polarization
state.

For the preparation of pure polarization states of qubits we selected a
spectral range about 1~nm. The initial state $\ket{H}$ passed through a phase
plate $h_d$ (436~$\mu$m) oriented at $0^\circ,45^\circ$. So the states under
consideration had the following form respectively:
\begin{equation}\label{eq:qubitstates}
\ket{\Psi_1}=\ket{H}, \ket{\Psi_2}=0.345\ket{H}-0.939i\ket{V}.
\end{equation}

The second requirement is achieved with a standard quantum tomography method
that used two achromatic quartz plates $h_2$=441~$\mu$m, $h_3$=313~$\mu$m
served for reconstruction of polarization states. These plates were oriented at
particular angles $\alpha,\beta$ (with respect to the vertical axis), so light
passing through the plates and following vertical polarizer was projected onto
the necessary set of states required by protocols J4 and R4. In the first
protocol J4, suggested in \cite{Kwiat}, projective measurements upon some
components of Stokes vector were performed:
$\ket{H},\ket{V},\ket{45^\circ}=\frac{1}{\sqrt2}(\ket{H}+\ket{V}),
\ket{R}=\frac{1}{\sqrt2}(\ket{H}+i\ket{V})$. Basically the projective
measurements can be chosen arbitrarily, in particular if the measured qubits
were projected on the states possessing tetrahedral symmetry then the protocol
transforms to R4. There are several works showing that due to the high symmetry
such protocol provides a better quality of reconstruction \cite{Rehacek, Ling,
Lingtwo}. Another protocol B9 exploits a single plate and fixed polarizer. The
corresponding measurements have been performed for each of the nine
orientations of the plate with a step of $20^\circ$. We have chosen the
``optimal'' thickness of the plate $h_3$=313~$\mu$m and achieved the better
condition number ($K=2.7$) for reconstruction of mixed states. Also for testing
theoretical predictions with this protocol we used other plates
$h_3$=824~$\mu$m and $h_3$=358~$\mu$m and corresponding condition numbers were
$K_{824}=14.9$ and $K_{358}=183.9$.

The third requirement was satisfied by using a single-photon detector based on
GaAs-based avalanche photodiode with a pigtailed input and an internal gate
shaper \cite{Molotkoff}. The total number of registered events was varied by
changing the gate rate at a fixed gate width (about 10~ns). This simply follows
from the fact that the fixed photon flux comes continuously and increasing
(decreasing) the gate width just increases (decreases) the probability to
register a photon.

\begin{figure}[t]
\includegraphics[width=0.95\columnwidth]{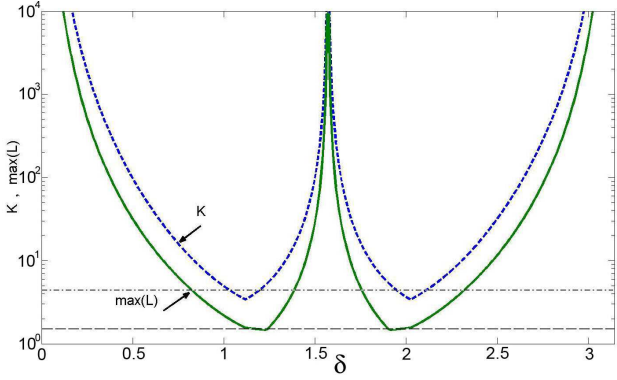}
\caption{(Color online) Calculated condition numbers K and maximum losses
$\max(L)$ on the Poincar\'e-Bloch sphere versus the optical thickness $\delta$
of a phase plate for the B9 protocol. The dashed and dash-dotted lines are for
the R4 and J4 protocols, respectively.} \label{f:K_L_distribution_qubit}
\end{figure}

The goal of our first experiment was the optimization of a quantum tomography
protocol for polarization qubits. As an example we selected protocol B9
implemented by means of a single phase plate and a fixed polarizer. Note that
the projection in the protocol B9 is performed onto non-fixed states, in
contrast to the protocols J4, K4, and depends on the parameters of the plate,
that is, on its optical thickness, $\delta=\pi h\Delta n/\lambda$, and the
orientation angle $\beta$, where $\Delta n$ is the birefringence of the plate
material at a given wavelength $\lambda$ and $h$ is its geometric thickness. It
is worth to mention that the parameters of this protocol can be chosen by doing
the experiment, depending on available resources: in some sense this choice can
be done in an optimal way, but using the plate with optimal condition number.
Figure~\ref{f:K_L_distribution_qubit} shows the calculated condition numbers
$K$ and maximum losses $L$ on the Poincar\'e-Bloch sphere as functions of the
optical thickness of the phase plate for the B9 protocol. The dependencies are
periodic with a period of $\pi$. Poor conditionality occurs at $\delta
=0,\frac{\pi}{2},\pi$. Both quantities $K$ and $L$ tend to infinity at these
particular points. For this protocol the best (lowest) achievable parameter $K$
takes the value 1.85. For example, to reach such a value one might choose
plates with optical thickness $\delta=0.356\pi$ or $\delta=0.644\pi$. The
corresponding condition numbers for protocols J4, R4 are
$K_{\text{J4}}\approx3.23$ $K_{\text{R4}}=\sqrt{3}\approx1.73$. Particularly
for our experiment we have chosen a set of following three plates:
$h_3$=313~$\mu$m (optimal), $h_3$=824~$\mu$m (medium), $h_3$=358~$\mu$m
(nonoptimal) with condition numbers $K=2.7$, $K=14.9$ and $K=183.9$. The
optimal value of maximum losses $L_{\text{opt}}=1.47$ occurs at the points
$\delta=0.391\pi, 0.609\pi$. The corresponding losses are somewhat lower than
those for the R4 protocol ($L=1.5$), and essentially lower than those for the
J4 protocol ($L=4.52$). Note that these curves on
Fig.~\ref{f:K_L_distribution_qubit} are similar in form, but their minima
correspond to slightly different values of $\delta$. However, the singular
values of $K$ and $L$ correspond to the same $\delta$ value. Thus, both
criteria allow separating the regions of the protocol parameters for which the
results would be unsatisfactory. Unfortunately, for a large dimension of the
Hilbert space for reconstructed states, the scanning of $L$ values is a
complicated computational problem. For this reason, to optimize such protocols,
we suggest to use the parameter $K$.

\begin{figure}[t]
\includegraphics[width=0.95\columnwidth]{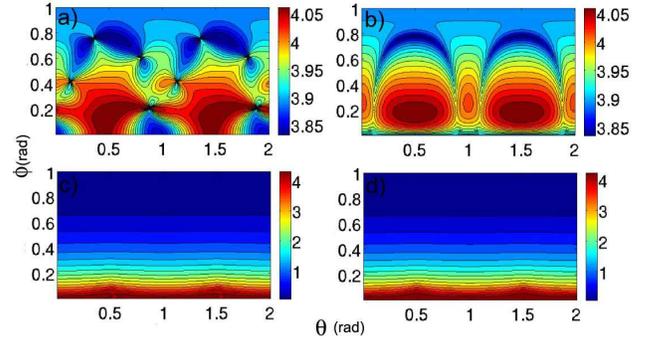}
\caption{(Color online) Difference between the (a) and (b) optimal and (c) and
(d) nonoptimal approach for  protocol B9. (a) and (c) and (b) and (d)
correspond to $20^\circ$ and $1^\circ$ steps of the plate orientation,
respectively.} \label{f:sphere_fidelity_distribution_B9}
\end{figure}

Figure~\ref{f:sphere_fidelity_distribution_B9} demonstrates the crucial
difference between the optimal and nonoptimal approach for the example of
considered protocol B9. Here the optimal protocol
[Figs.~\ref{f:sphere_fidelity_distribution_B9}(a) and
\ref{f:sphere_fidelity_distribution_B9}(b)] corresponds to the plate with
thickness $h$=293.8~$\mu$m and condition number $K=2.2548$. This protocol
provides the  minimum of maximally possible losses (minimax). The nonoptimal
protocol [Figs.~\ref{f:sphere_fidelity_distribution_B9}(c) and
\ref{f:sphere_fidelity_distribution_B9}(d)] corresponds to the plate with
thickness $h$=358~$\mu$m and condition number $K=183.9$. The sample size is
$10^4$. Figures~\ref{f:sphere_fidelity_distribution_B9}(a) and
\ref{f:sphere_fidelity_distribution_B9}(b), which correspond to the optimal
protocol, describe the case of guaranteed number of nines in Fidelity being not
less than $z_{\text{min}}=3.83$ (so that the mean fidelity (\ref{eq:aver}) for
all states on Bloch sphere can not be lower). At the same time,
Figs.~\ref{f:sphere_fidelity_distribution_B9}(c) and
\ref{f:sphere_fidelity_distribution_B9}(d), which correspond to the nonoptimal
protocol, stand for a very low level of fidelity ($z_{\text{min}}=0.065$),
which means that almost none of the states on Bloch sphere can be reconstructed
using the protocol. Figures~\ref{f:sphere_fidelity_distribution_B9}(b) and
\ref{f:sphere_fidelity_distribution_B9}(d) are different from
Figs.~\ref{f:sphere_fidelity_distribution_B9}(a) and
\ref{f:sphere_fidelity_distribution_B9}(c) by the step of the plate orientation
[20 degrees for Fig.\ref{f:sphere_fidelity_distribution_B9}(a) and
\ref{f:sphere_fidelity_distribution_B9}(c) and 1 degree for
Figs.~\ref{f:sphere_fidelity_distribution_B9}(b) and
\ref{f:sphere_fidelity_distribution_B9}(d), respectively], so that the number
of rows in the protocol grows from 9 to 180. We see that
Fig.~\ref{f:sphere_fidelity_distribution_B9}(b) defines a more symmetric
distribution of precision compared to
Fig.~\ref{f:sphere_fidelity_distribution_B9}(a), even though the range of
values [$z_{\text{min}}$, $z_{\text{max}}$] barely changes. A comparison of
Fig.~\ref{f:sphere_fidelity_distribution_B9}(c) with
Fig.~\ref{f:sphere_fidelity_distribution_B9}(d) shows that in the nonoptimal
case increasing the number of projections little affects the distribution of
Fidelity on Bloch sphere.

\begin{figure*}[t]
\includegraphics[width=2\columnwidth]{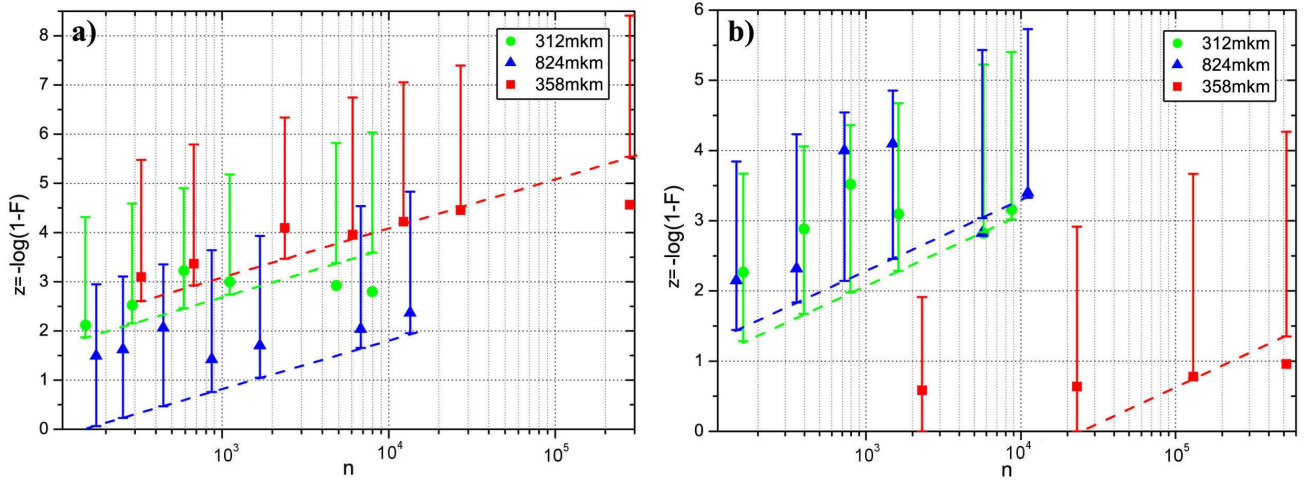}
\caption{(Color online) Reconstruction of pure qubit states by protocol B9 at
different thicknesses of the plate. Vertical bars show 1\% and 99\% quantiles
for fidelity distributions. Dotted lines connecting lower bar ends point out
critical significant levels.} \label{f:experiment_qubit_pure}
\end{figure*}

For protocol B9 the statistical reconstruction of the prepared states
(\ref{eq:qubitstates}) has been performed at given sample sizes. One of the
considered states $\ket{\Psi_1}$ belongs to an area of small losses for all
three plates, and the second state $\ket{\Psi_2}$ was changed out of this area.
As an example, we present Fig.~\ref{f:experiment_qubit_pure} that shows the
calculated widths of fidelity distributions at $1\%$ and $99\%$ quantiles as
well as the experimentally reconstructed values for the states
(\ref{eq:qubitstates}). Two polarization states (\ref{eq:qubitstates}) were
measured and reconstructed using three phase plates chosen above. Experiments
were performed with different sample sizes; that is, the total number of the
pulses coming from single-photon detector in a fixed time was varied and served
as a parameter of the problem.

The approach described above provides the ideal accuracy level for quantum
state reconstruction. It means that the fluctuations of the estimated quantum
states cannot lead to uncertainties smaller than this limit. The presence of
instrumental errors and uncertainties makes this level to be exceeded. Indeed,
Fig.~\ref{f:experiment_qubit_pure} shows that, above some sample size, the
experimental value of fidelity falls out the theoretical uncertainty boundary
shown as dotted lines (corresponding to $1\%$ significance level which
characterizes a given protocol). This happens since instrumental uncertainties
prevail over the statistical ones and indicates that either state preparation
stage or measurement procedure were not performed accurately enough. It is
clearly seen that both theoretical distributions and experimental points for
protocol with nonoptimal choice of plates strongly depend on an initial state.
The quality of reconstruction (fidelity) is varied in the range 0.6030--0.9999
for both reconstructed states for the protocol with the plate $h_3$=358~$\mu$m
(nonoptimal) and in the range 0.9950--0.9997 for the plate $h_3$=313~$\mu$m
(optimal). It means that if  the set of initial states is known (as it happens
in quantum process tomography) one can choose the parameters of the protocol
that are not optimal for an arbitrary state, but provide the highest accuracy
for the selected set of states. In both cases, when the input states are
unknown, like in quantum state tomography, it is better to use optimal choice
of the plates with minimal condition number, that will ensure high accuracy of
the reconstruction of arbitrary states.

\textit{Mixed States: modeling and experiment.} The goal of the second
experiment with qubits was the generalization of the developed approach to the
family of mixed states of polarization qubits \cite{general-appr}. As an
example we have tested three protocols again, namely J4, R4, and B36. The
protocol B36 has been considered at optimal parameters (phase plate
$h_3$=313~$\mu$m). This protocol is similar to the one described above (B9),
but measurements were performed at 36 consistent orientations of the phase
plate ($0^\circ$--$360^\circ$, step $10^\circ$) instead of 9.

As we have pointed out before, the condition numbers $K$ for R4 and J4
protocols take the following values: $K_{\text{R4}} = \sqrt{3}\approx1.73$,
$K_{\text{J4}} \approx 3.23$ while for B36 it becomes $K_{\text{B36}}\approx
2.7$. Thus, we expect that the symmetrical protocol $R4$ provides with better
state reconstruction quality. We have checked this statement with numerical
simulations of each protocol applied to qubit states with various degree of
mixture depending on the sample size.

For the preparation of mixed states we started from a pure state
$\ket{\Psi_1}=\ket{H}$ which passed through one or two thick birefringent
plates, oriented at $45^\circ$, where partial decoherence between vertical and
horizontal basic polarizations took place. For that purpose we used a quartz
plate with a thickness of $h_0$=10~mm and additional tiff plate with a
thickness of $h_1$=4~mm. Changing the width of the spectrum it is possible to
pass from completely coherent (pure) case to a very narrow spectral line to a
completely incoherent (i.e. mixed in polarization), when the difference between
the optical lengths for two orthogonal polarizations exceeds the coherence
length of the radiation under consideration:
\begin{equation}
h_0(n_{\bot}-n_{\|})_{\text{quartz}}+h_1(n_{\bot}-n_{\|})_{\text{tiff}})\gg
l_{coh},
\end{equation}
where
$$
l_{coh}\approx\frac{\lambda^2}{\triangle\lambda}\approx100~\mu\text{m}.
$$
Thus, at the output of plates $h_0,h_1$ components with orthogonal
polarizations evolve with a random phase, which leads to a mixed state.

For determining the accuracy of quantum tomography and fidelity, we
need to calculate the density matrix of a prepared mixed state. For
this purpose we divide the frequency spectrum of the radiation in
small parts and represent the polarization state of a qubit as a
superposition of states corresponding to different frequencies in
the spectrum:
\begin{equation}\label{f:qubit_state}
\begin{array}{cc}
\ket{\Psi}=\sum_k a_k \ket{\Phi(\omega_k)},\\
\ket{\Phi(\omega_k)}=c_1(\omega_k)\ket{H}+c_2(\omega_k)\ket{V},
\end{array}
\end{equation}
where amplitudes $a_k$ are defined by the spectrum shape. The density matrix of
the state before transformation (\ref{f:qubit_state}) has the form:
\begin{equation}\label{f:qubitmatrix}
\rho^{\text{in}}=\ketbra{\Psi}{\Psi}=\sum_{k,j} a_{k}a_{j}^{\ast}
\ketbra{\Phi(\omega_k)}{\Phi(\omega_j)}
\end{equation}

The next step of modeling includes the calculation of phase plates action on the qubit state.
The unitary transformation on state (\ref{f:qubitmatrix}) is given by matrix
\begin{equation}\label{f:matrixG}
G(\omega_{k})=\left(
                \begin{array}{cc}
                  t_{k} & r_{k} \\
                  -r_{k}^\ast & t_{k}^\ast \\
                \end{array}
              \right)
\end{equation}
where
\begin{equation}
\begin{array}{cc}
t_{k} = \cos \delta_{k} + i\sin \delta_{k} \cos 2\alpha,\\
\quad r_{k} = i\sin \delta_{k} \sin 2\alpha , \delta_{k} =
{\pi (n_{o}^{k} - n_{e}^{k}) h/ \lambda_{k} }.
\end{array}
\end{equation}

Here $t_k$ and $r_k$ are the amplitude transmission and reflection coefficients
of the wave plate at fixed frequency, $\delta_k$ is its optical thickness, $h$
is the geometrical thickness, $\alpha$ is the orientation angle between the
optical axis of the phase plate and vertical direction. The measurement part of
the experimental setup does not distinguish frequency modes, so the theoretical
polarization mixed state will be described by a reduced density matrix:
\begin{equation}\label{f:ro}
\rho=\sum_{k}|a_{k}|^2G(\omega_k)\ketbra{\Phi(\omega_k)}{\Phi(\omega_k)}G^{+}(\omega_{k}).
\end{equation}

\begin{figure*}[t]
\includegraphics[width=1.5\columnwidth]{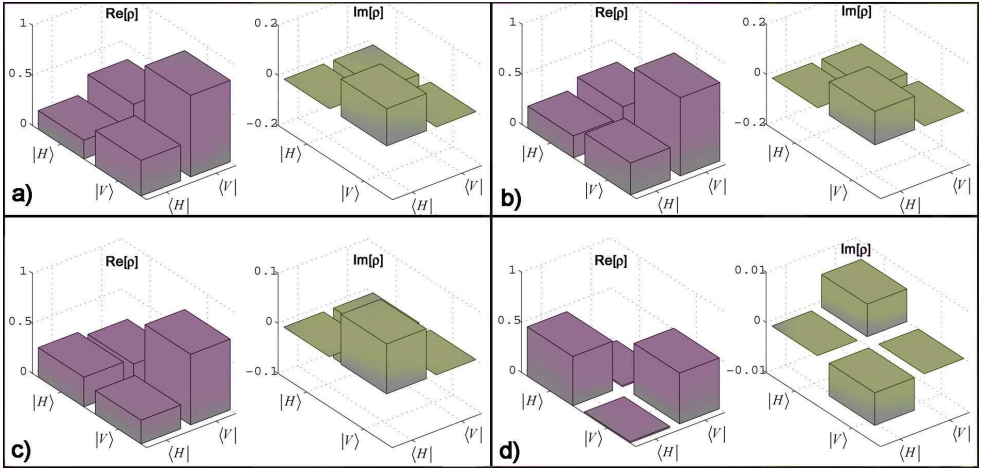}
\caption{(Color online) Theoretical distribution of the real and imaginary
parts of the density matrix for states with varying degrees of purity. The
spectral widths of the spectrum are (a) 1.6~nm, (b) 7~nm, (c) 13~nm.
Distribution (d), for 22~nm, corresponds almost to a completely mixed state.}
\label{f:qubit_mixed_theory}
\end{figure*}

The corresponding density matrixes are obtained by integrating formula
(\ref{f:ro}) with the distribution of weights $|a_{k}|^{2}$ as a function
$\hbox{sinc}^2(x)$. Figure~\ref{f:qubit_mixed_theory} shows the graphical
representations of the real and imaginary parts of the theoretical qubit
density matrices at different widths of the spectrum: 1.6~nm, 7~nm, 13~nm,
22~nm. When increasing the width of the spectrum the non-diagonal components,
responsible for the correlation, decay and the state becomes completely mixed.

\begin{figure*}
\includegraphics[width=2\columnwidth]{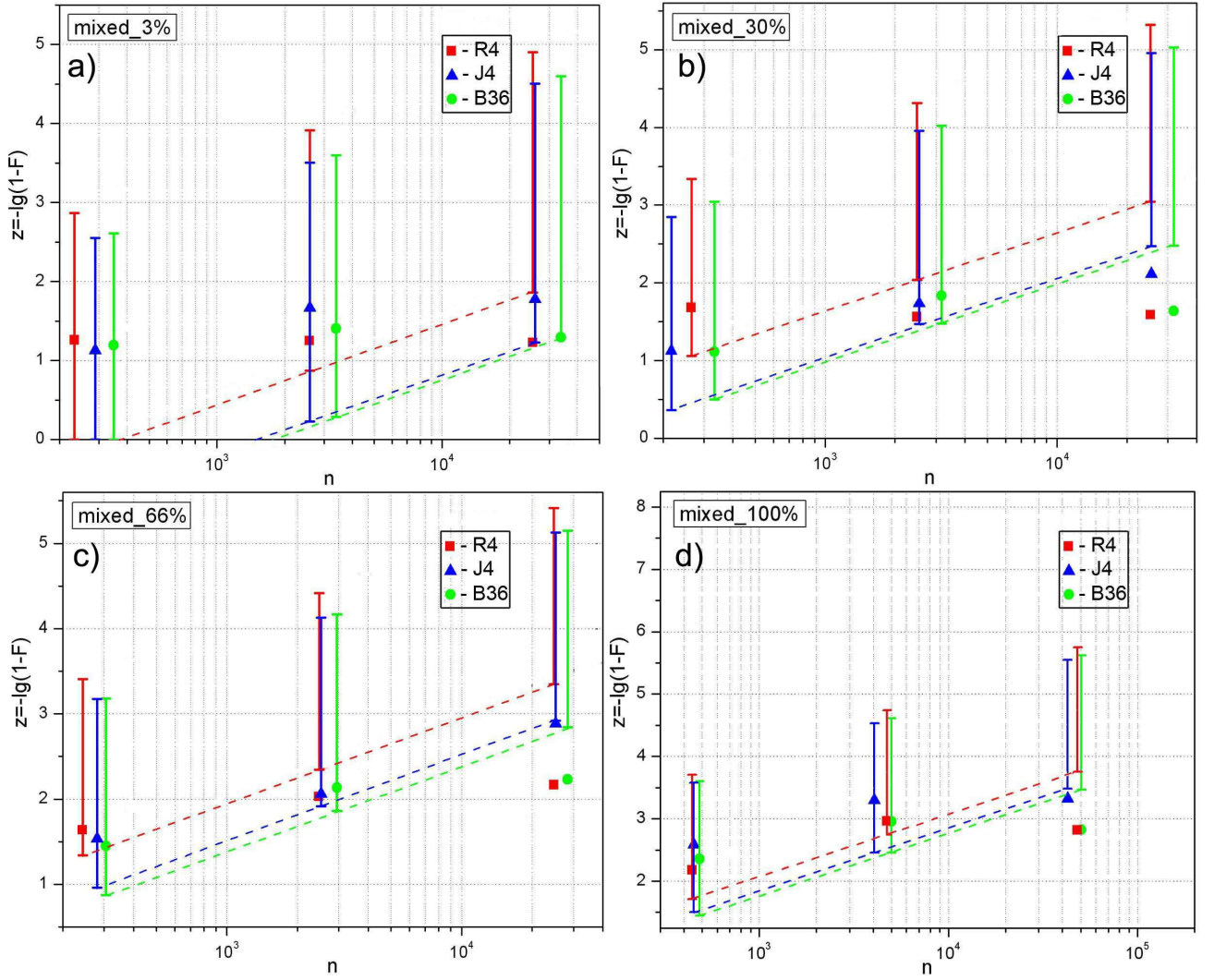}
\caption{(Color online)Reconstruction of qubits states with various degree of
mixture. Vertical bars show 1\% and 99\% quantiles for fidelity distributions.
Dotted lines connecting lower bar ends point out critical significant levels.
(a) mixed 3\%, (b) mixed 30\%, (c) mixed 66\%, and (d) mixed 100\%.}
\label{f:experiment_qubit_mixed}
\end{figure*}

In the experiment we have prepared four polarization states of qubit with
following degrees of mixture: 3\% (1.6~nm), 30\% (7~nm), 66\% (13~nm), 100\%
(22~nm), where the state purity is analyzed by calculating the state entropy
defined as $S=-\sum_{n=1}^2\lambda_n\log_2\lambda_n$, $\lambda_n$ being the
eigenvalues of the density matrix $\rho$. For each protocol various
measurements at different sample sizes were performed. The results are shown in
Fig.~\ref{f:experiment_qubit_mixed}.

Figure~\ref{f:experiment_qubit_mixed} presents the theoretical distributions of
fidelity at 1\% and 99\% quantiles for each considered protocol. Experimental
values are indicated by points. The dashed lines connecting the 1\% quantiles
show a theoretical lower boundary of the fidelity distribution and depend on
the type of protocol: the lower line, the greater condition number and the
function of losses. Fig.~\ref{f:experiment_qubit_mixed} demonstrates that for a
small sample size statistical uncertainties dominate over the instrumental
ones. Starting from sample size about $(2-3)\times10^3$ the experimental points
fall out the theoretical uncertainty boundary and the protocol achieves a
coherent sample size. This means that instrumental uncertainties (the accuracy
of setting angles, thicknesses of phase plates, etc.) prevail over the level of
statistical errors. From this point a further increase of sample size does not
improve the quality of the reconstruction of the quantum state. However, a
comparison of experimental results with the theoretical distribution serves as
effective method for setup adjusting, stability of the preparation/measurement
systems, etc.

Furthermore, from Fig.~\ref{f:experiment_qubit_mixed} one can evince that the
accuracy of the reconstruction increases with the degree of mixture: for the
states close to a pure ones (3\% mixture) the fidelity achieves an average
value 0.93--0.94, while the best accuracy is achieved for totaly mixed state
corresponding to the center of the Bloch sphere, $F>0.99$. The theoretical
estimates are consistent with both numerical and real experiments.

\textit{Ququarts: pure and mixed states, preparation and measurement.} For a
further verification of  our approach we prepared also a family of quqarts
biphoton polarization states \cite{se}, which can be easily converted into
either entangled (in polarization) or factorized states, both pure and mixed.
\begin{equation}\label{eq:family}
\begin{array}{cc}
\rho =(1-p)\ketbra{\Psi_{\text{pure}}}{\Psi_{\text{pure}}}
+\frac{p}{2}(\ketbra{H_1H_2}{H_1H_2} \\
+\ketbra{V_1V_2}{V_1V_2}),
\end{array}
\end{equation}
where
\begin{equation}\label{eq:Psi}
\ket{\Psi_{\text{pure}}}= c_1\ket{H_1H_2} +c_2e^{i\varphi}\ket{V_1V_2}
\end{equation}
with real amplitudes $c_1$ and $c_2$ and relative phase shift $\varphi$. The
coefficient $p$ defines the degree of mixture.

\begin{figure}
\includegraphics[width=1\columnwidth]{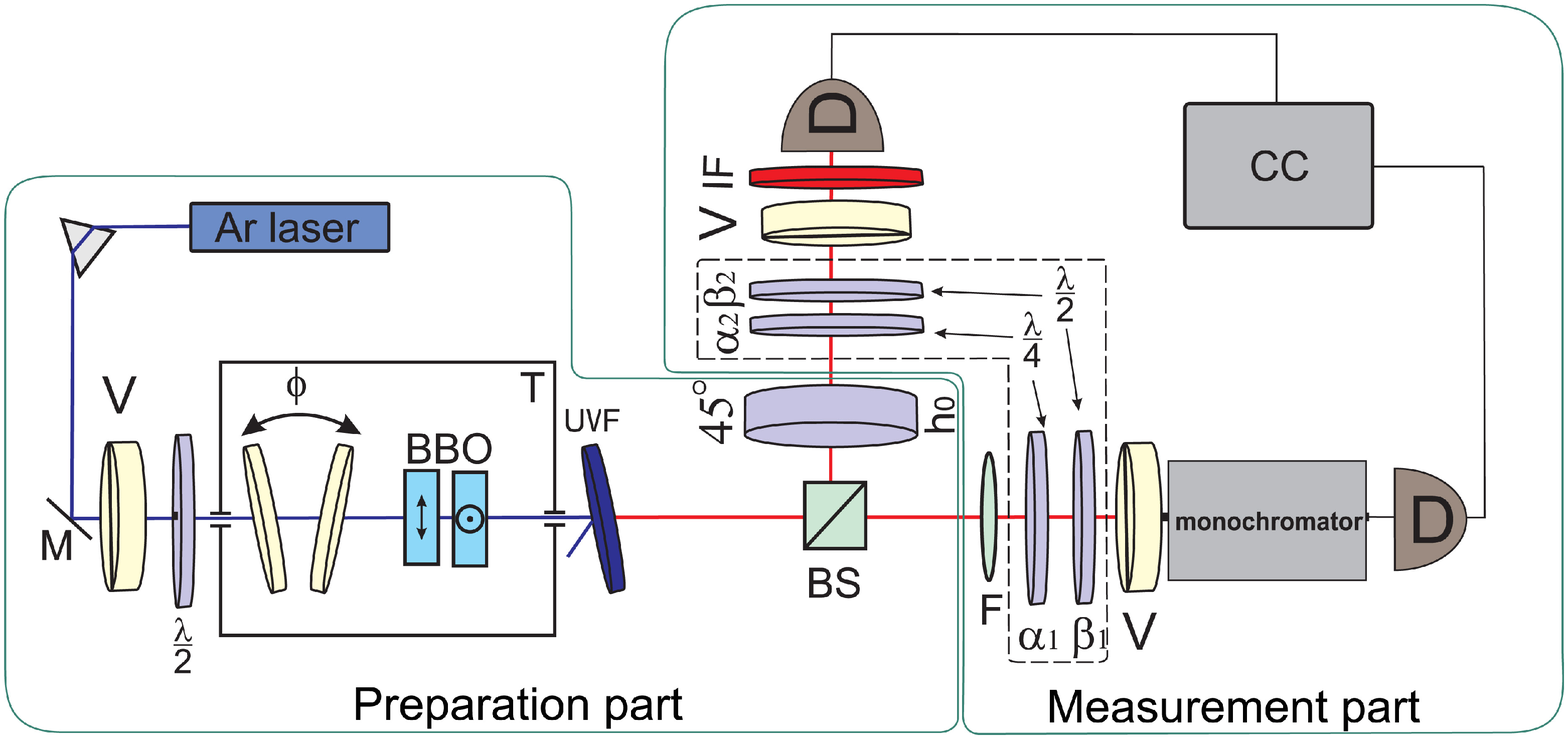}
\caption{(Color online) Experimental setup for different tomographic
reconstructions of photon pairs with variable polarization entanglement and
degree of mixture. Ar laser: argon laser with wavelength 351~nm, M: mirror, V:
vertical oriented Glan-Thompson prism, BBO: nonlinear Barium Borate crystals,
UVF: ultraviolet filter, $\lambda_p/2$, $\lambda_p/4$: half-wave, quarter-wave
plates, L: lens with focus 20~cm, BS: beamsplitter, IF: interfilter, D:
avalanche photodetectors, CC: coincidence circuit.} \label{f:setup_ququart}
\end{figure}

The set-up is schematically depicted in Fig.~\ref{f:setup_ququart}. For
generation of pure biphoton-based polarization ququarts ($p=0$) we used a set
of two orthogonally oriented type-I BBO crystals (1~mm), cut for collinear,
frequency non-degenerate phase-matching around the central wavelength of
702~nm. The crystals were pumped by a 600~mW cw-argon laser operating at
351~nm. The Glan-Thompson prism (V) at vertical polarization and the half-wave
plate ($\lambda_p/2$) placed in front of crystals allowed rotating the
polarization of the pump by the angle $\phi$, which controlled the real
amplitudes $c_1$ and $c_2$  in (\ref{eq:Psi}). A set of quartz plates $QP$
introduced the relative phase shift $\varphi$ between the horizontal and the
vertical components of the UV pump. If $\phi=0$ we prepared the state
$\ket{\Psi}=\ket{V_1V_2}$, if $\phi=22.5^\circ$ and $\varphi=3\pi/2$ then the
state transformed to $\ket{\Psi}=\frac{1}{\sqrt{2}} \left(\ket{H_1H_2} -
\ket{V_1V_2} \right)\equiv \ket{\Phi^{-}}$. To maintain stable phase-matching
conditions, BBO crystals and QP were placed in a closed box heated at fixed
temperature. The lens $L$ coupled SPDC light into the monochromator M (with
4~nm resolution), set to transmit ``idler'' photons at 710~nm. The conjugate
``signal'' wavelength 694~nm was selected automatically by means of the
coincidence scheme.

In order to prepare a mixed state it is necessary to introduce quantum
distinguishability among the biphoton ququart basis states in a controllable
manner. For the preparation of states with various degree of mixture the
double-crystal scheme was complemented by a quartz plate with thickness $h_0$,
which was putted in the reflected arm of Brown-Twiss scheme. A thick quartz
plate with vertically oriented optical axis introduced a delay between
vertically and horizontally polarized photons that led to their temporal
distinguishability, and hence, to the emergence of a mixture. Changing the
width of the frequency spectrum, which, in our case, was determined by the
width of the monochromator slit, or the plate thickness it was possible to
change the visibility of polarization interference. Our goal was the
preparation and reconstruction (by several protocol, including non-optimal
ones) of a set of states with different degree of entanglement and set of mixed
two-qubit states.

The reconstruction part for the ququart is shown in Fig.~\ref{f:setup_ququart}.
First, the photon pair that forms the biphoton-ququart was split into two
spatial modes by using a beam splitter BS. A photon in each arm, then,
underwent a polarization state transformation with the use of a couple of
zero-order wave plates ($\lambda/2$,$\lambda/4$) for protocols. Two Si-APD's
linked to a coincidence scheme with 1.5~ns time window were used as single
photon detectors. Registering the coincidence rate for different projections,
that were realized by the half- and quarter-plates and a fixed analyzer, was
possible to reconstruct the polarization state of ququarts by protocols R16,
J16, described above. This scheme also allowed reconstructing any arbitrary
polarization ququart state by protocol B144. For the reconstruction of ququarts
pure states by protocol B144 we used two retardation plates Wp1, Wp2 placed
before beam splitter and deleted half-wave and quarter-wave plates.
Unfortunately, this method can not be applied for the reconstruction of mixed
states of ququarts. This is due to the feature of preparation for mixed states:
all projection measurements must be done after state preparation that is
violated for such configuration of the tomography protocol. To solve this
problem, we have used an equivalent scheme. In each arm we have established a
couple of identical plates Wp1, Wp2 instead of half-wave, quarter-wave plates
and rotated them synchronously. This scheme is completely equivalent to the
standard protocol B144, but slightly more complicated in realization. By
analogy to protocols B9 and B36 with single retardant plate we can choose the
optimal parameters for protocol B144.

Let us analyze how the choice of plates thicknesses affects the quality of
state reconstruction.

\begin{figure}
\includegraphics[width=1\columnwidth]{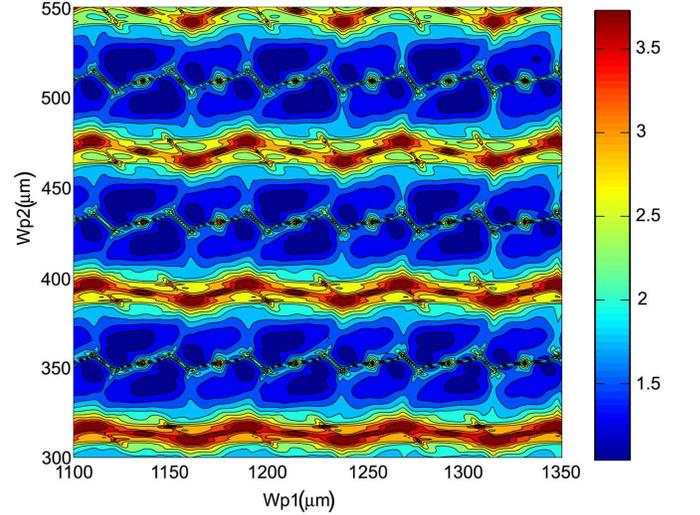}
\caption{(Color online) Dependence of  $\log(K)$ on plates thicknesses for
protocol B144.} \label{f:2Dratio}
\end{figure}

Figure~\ref{f:2Dratio} presents the calculated condition number $K$ in common
logarithmic scale for the set of thicknesses of the first and second quartz
retardant plates. Specifically the first plate Wp1 was varied within the limits
from 1.100 to 1.350~mm with a step 0.001~mm and a second plate, Wp2, was varied
within the limits from 0.300 to 0.550~mm with the same step. Here the
thicknesses of the first and second quarts retardant plates are shown along the
axis $x, y$, correspondingly. Blue color indicates areas with low value of
condition number $K$, while the red one indicates areas with high value of $K$.
It is obvious, that the arbitrary choice of plates most likely will be wrong
from the point of view of completeness of matrix $B$ and protocol optimality.
In other words this figure serves as some sort of a ``navigation map'' for the
measurement protocol. So to achieve good quality of the measurement one should
select the plates thicknesses in blue areas and avoid red areas. For the
present protocol the minimum achievable $K$ is 11.35. For example, to reach
such a value one can choose plates with thickness Wp1=1.207~mm and
Wp2=0.520~mm. In our specific experimental configuration we have selected the
plates Wp1=1.303~mm and Wp2=0.440~mm. Such thicknesses of the plates for the
protocol B144 were chosen to show the difference in the quality of
reconstruction.

For all three protocols used in the experiment we have calculated the condition
numbers $K$, which assume the following values: $K_{\text{R16}}=3$,
$K_{\text{J16}}\approx 10$, $K_{\text{B144}}\approx 60$.  Thus, we expect that
the symmetrical protocol R16 provides with better state reconstruction quality
\cite{Burgh} and the protocol B144 provides the worst one. We have checked this
statement with numerical simulations of each protocol applied to different
two-qubit states depending on the sample size.

\begin{figure}[t]
\includegraphics[width=1\columnwidth]{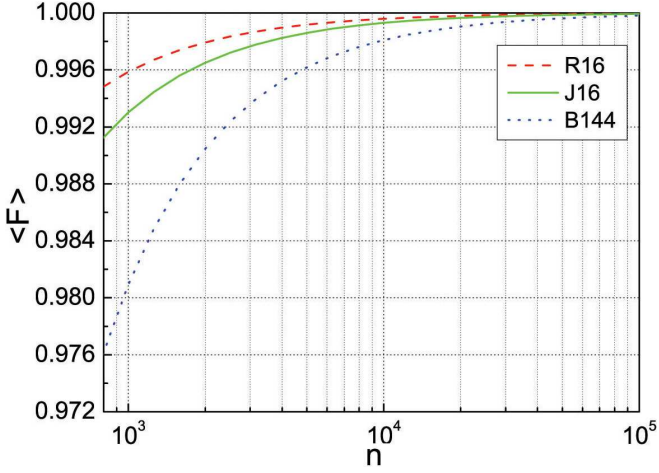}
\caption{(Color online) Dependence of the average fidelity on number of
registered events forming the sample for Bell state reconstructed by protocols
J16, K16 and B144.} \label{f:F_N}
\end{figure}

\begin{figure}[b]
\includegraphics[width=1\columnwidth]{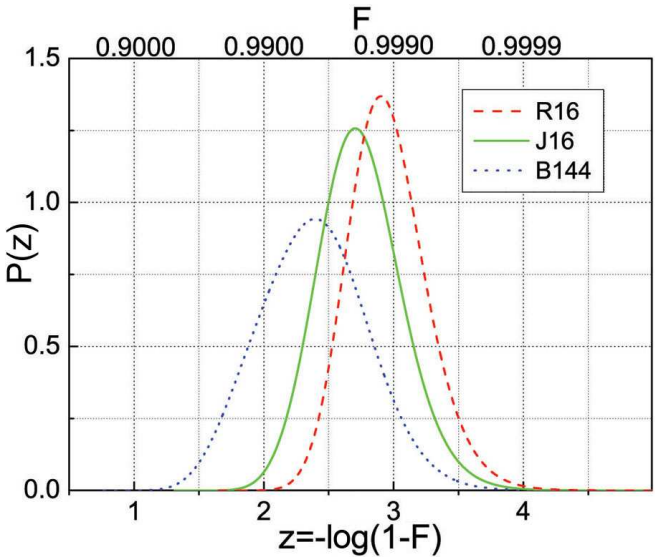}
\caption{(Color online) Density distribution  of the scaled fidelity $z$ (lower
abscissa) at $n=3\times 10^{3}$ for Bell state reconstructed by protocols J16,
K16 and B144. Upper abscissa presents regular fidelity.} \label{f:losses}
\end{figure}

As an example, let us consider the numerical reconstruction of the Bell state
$\ket{\Phi^{-}}=\frac{1}{\sqrt{2}} \left(\ket{H_1H_2}-\ket{V_1V_2}\right)$.
Figure~\ref{f:F_N} shows average fidelities, calculated according to
(\ref{eq:average_losses}) as functions of a sample size for each protocol. It
turns out that the difference between curves disappears at sufficiently large
sample size ($10^5$)\cite{instrumental}, but for the same quality of state
reconstruction the correct choice of protocol allows using a smaller set of
statistical data, i.e. finally reduces total acquisition time.
Figure~\ref{f:F_N} shows that protocols are ranged in accuracy as following:
R16, J16, and B144 in complete agreement with the range given by condition
number $K$. Figure~\ref{f:losses} presents accuracy distributions calculated
according to (\ref{eq:average_losses}) for the sample size $3\times10^3$.

It is clearly seen that the density distribution for the R16 protocol is
narrower than the one for J16 and B144 and localizes in the region of lower
losses or higher fidelities. The distribution for B144 is broader and lower in
comparison with R16 and J16. Obviously, the narrower distribution of fidelity
indicates a better reconstruction quality in the sense that the reconstruction
procedure returns a better-defined state. Thus, Figure~\ref{f:losses} confirms
our expectation based on estimation of condition number $K$: R16 achieves the
best results.

\begin{figure*}[t]
\includegraphics[width=2\columnwidth]{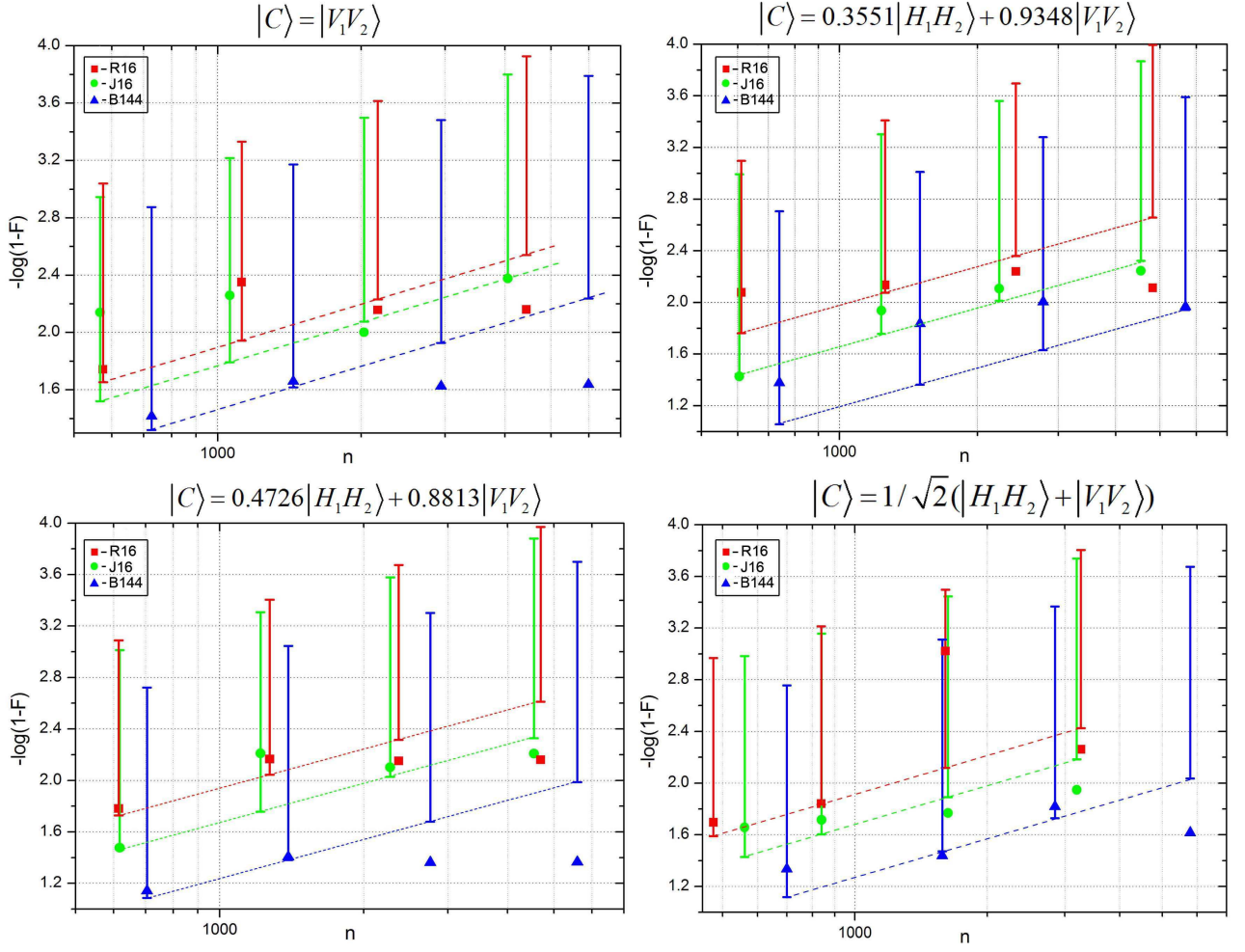}
\caption{(Color online) Reconstruction of ququart states with various degree of
entanglement. Vertical bars show 1\% and 99\% quantiles for fidelity
distributions. Dotted lines connecting lower bar ends point out critical
significant levels.} \label{f:experiment_ququarts_pure_entang}
\end{figure*}

Let us consider now the experimental results. As an example,
Fig.~\ref{f:experiment_ququarts_pure_entang} shows both experimental points and
theoretical distributions for the set of pure ququart states with various
degree of entanglement: $C_{\varphi_1}=0$, $C_{\varphi_2}=0.66$,
$C_{\varphi_3}=0.83$, $C_{\varphi_4}=1$ correspondingly. Here $C_{\varphi_i}$
is quantity concurrence, which is defined as $C=2|c_1c_4-c_2c_3|$ and $C=0$ for
a separable state and $C=1$ for a maximally entangled state.

\begin{figure*}[t]
\includegraphics[width=2\columnwidth]{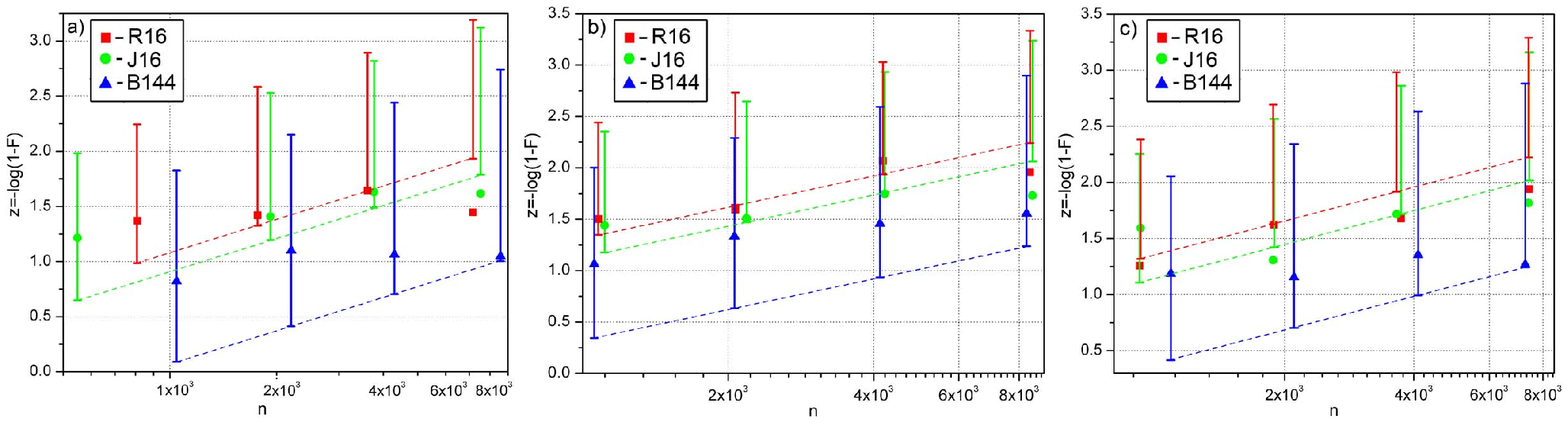}
\caption{(Color online) Reconstruction of ququart states with various degree of
mixture. Vertical bars show 1\% and 99\% quantiles for fidelity distributions.
Dotted lines connecting lower bar ends point out critical significant levels.
(a) $p=0.3$, (b) $p=0.7$, and (c) $p=1$.} \label{f:ququarts_mixed}
\end{figure*}

Figure~\ref{f:ququarts_mixed} presents experimental points and theoretical
distribution for the set of mixed ququart states with various degree of
mixture: $p=1,~0.3,~0.7$ ($p=1$ corresponds to totaly mixed state, $p=0.3$
means that state contains 30\% of mixture.)

On both figures the agreement between the experimental results and \textit{a
priori} calculated expectations is noteworthy.

Figure~\ref{f:ququarts_mixed} shows that the accuracy of reconstruction
increases with the degree of mixture. However from a statistical point of view
there is a problem. A small weight of the mixed states, adding to the original
pure state, leads to a small change in the statistical data while the number of
parameters for ququart state increases from 6 to 15. As a result the amount of
new information that arises in quantum measurements is not sufficient to
achieve an adequate estimation of all the parameters.

\section{Discussion}\label{sec:Discussion}

Many remarkable works in quantum state tomography developed different
theoretical approaches. In the present section we shall briefly discuss the
features and advantages of our approach in this framework.

One of the main issues in quantum tomography is the choice of adequate
parameterization for quantum states. Bloch sphere representation is one option.
In this representation density matrix $\rho$ is defined in Hilbert space of
dimension $s$ by the following equation \cite{Kwiat,Nunn,Paris-Reh}
\begin{equation}\label{eq:rho_discuss}
 \rho =\frac{1}{s}E+\nu_j\sigma_j,
\end{equation}
where $E$ is the $s\times s$ identity matrix, $\nu_j$ is a vector in real
Euclid space of size $s^2-1$, and $j$ changes from 1 to $s^2-1$. Respective
Hermitian basis matrices $\sigma_j$ meet the following conditions:
\begin{equation}
\Tr(\sigma_j)=0,\,\,\,\,\, \Tr(\sigma_j\sigma_k)=\delta_{jk}.
\end{equation}
Note, that for single-qubit states when $s=2$, matrices $\sigma_j$ coincide with
ordinary Pauli matrices (multiplied by $\frac{1}{\sqrt{2}}$). The radius of 3D
Bloch sphere in this representation is also equal to $\frac{1}{\sqrt{2}}$.

One may easily calculate Bloch vectors $\nu_j$ from the known density matrix $\rho$:
\begin{equation}
\nu_j=\Tr(\rho \sigma_j).
\end{equation}
However, Bloch parameterization has a significant drawback. For every real
vector $\nu$ Eq. (\ref{eq:rho_discuss}) guarantees that the density matrix
$\rho$ is Hermitian, but does not guarantee that it is positively defined. As a
result, we can obtain a non-physical density matrix using Eq.
(\ref{eq:rho_discuss}). Separation of admissible density matrices from those
that can not physically exist is a non-trivial task for all dimensions higher
than two.  In particular, for $s>2$ Bloch representation is practically useless
for description of matrices of non-full rank, for which the number of degrees
of freedom of quantum state (\ref{eq:degrees}) is less than dimension of Bloch
vector $s^2-1$. This drawback of Bloch parameterization complicates the use of
Fisher information matrix and calculation of statistical estimates' precision
in general.

An adequate and convenient procedure of parameterization is important for
numerical procedures of calculating density matrix from experimental data
because for every iteration the approximate solution must lie in the range of
physical density matrices only. For instance, such issues may arise for
problems of estimation of quantum state by maximum likelihood method.

The problem of estimating a density matrix by maximum likelihood method was
considered in \cite{Hradil,T,Zambra,Allevi}. The respective likelihood equation
has the following form \cite{Hradil,Paris-Reh,Hradil-Summ}:
\begin{equation}
 R\rho =\rho,
\end{equation}
where $R$ is some operator.

The iteration procedure from $j$th to $j+1$th step for this equation has the
following form: $\rho^{j+1}=R\rho^{j}$. Such procedure, however, does not even
provide Hermitian property for density matrix.

Another procedure that is widely used \cite{Paris-Reh}
$\rho^{j+1}=\frac{1}{2}(R\rho^{j}+\rho^{j}R)$ provides the Hermitian property
but does not guarantee positive definiteness in general. Fortunately there are
known robust, well-converging procedures providing with both Hermiticity and
positive definiteness simultaneously \cite{Lv}.

In this paper we consider a natural approach for quantum theory that is based
on representing the quantum state as a mix of pure states
\cite{Bogdanov,Bogdanovtwo,JETPLBogd}. Consider transformation of the density
matrix to diagonal form.
\begin{equation}
 \rho=UDU^\dag.
\end{equation}
Here $U$ is an unitary matrix (its columns define eigenvectors of density
matrix), $D$- is a diagonal non-negative matrix (its diagonal is formed by
eigenvalues of density matrix that we shall put in decreasing order). Let the
rank of density matrix be $r$ ($1\leq r \leq s$). We shall eliminate all
non-zero rows and columns of matrix $D$  and shall leave only the first $r$
columns in matrix $U$. Then the density matrix could be represented in the
following compact form.
\begin{equation}\label{eq:LL}
 \rho =LL^\dag,
\end{equation}
where $L=UD^{\frac{1}{2}}$.

Here $L$ is a complex matrix of dimension $s\times r$ that defines the actual
parameterization of density matrix. It represents a purified state vector
(probability amplitudes). For representation as a column vector, one should
simply place the second column under the first one and so on. Note that the
purified state is defined ambiguously, because the density matrix does not
change during transformation
\begin{equation}\label{eq:L_implication}
 L\rightarrow L^{\prime}=LV,
\end{equation}
where $V$- is an arbitrary unitary matrix of size $r\times r$.

It is crucial that the ambiguity (\ref{eq:L_implication}) does not affect the
procedure of statistical reconstruction of density matrix, because all possible
purified state vectors define the same density matrix.

We can choose the unitary transformation $V$ such that in matrix $L^{\prime}$
all elements on the principal diagonal will have real strictly positive values,
while elements higher the principal diagonal shall be equal to zero.

For a density matrix of full rank, such matrix $L^{\prime}$ shall form a lower
triangular matrix. In mathematics such decomposition is usually called Cholesky
decomposition. Representation of density matrix in such form was originally
studied in \cite{Banaszektwo}. Note that from the point of view of physics,
Cholesky representation is just one of the possible ways of recording a
purified quantum state.

A purified state vector represents the most simple and most natural in view of
physics form of parameterization of density matrix. It is important that
parameterization based on purification of quantum state radically simplifies
the theory of statistical estimates.

For example, the precision of estimates by maximum likelihood method is defined
by matrix of full information that is analogous to Fishers information matrix
in relation to estimation of quantum state vector
\cite{Bogdanov,Bogdanovtwo,JETPLBogd}.

The matrix of full information is defined by the following equation:
\begin{equation}\label{eq:H_matrix}
 H=2\sum_j\frac{t_j(\Lambda_jc)(\Lambda_jc)^+}{\lambda_j}.
\end{equation}
The object is defined in real Euclidian space of dimension $2rs$. To obtain
state vector $c$ in this representation, one should place the imaginary part of
purified state vector under its real part. Intensity matrices $\Lambda_j$
represent expansion of matrices (\ref{eq:lambdamix}) to the considered Euclid
space. The sum in Eq. (\ref{eq:H_matrix}) is taken for all measurement protocol
rows. Matrix $H$ becomes a real symmetric matrix of dimension $2rs\times 2rs$.

If the quantum measurement protocol is complete, then for the information
matrix $H$ $\nu_H=(2s-r)r$ out of $2rs$ eigenvalues are strictly positive,
while the other $r^2$ are equal to zero. One may formulate a universal
statistical distribution (\ref{eq:average_losses}) based on information matrix
$H$ that will describe precision of statistical reconstruction of quantum
states.

Eigenvectors of matrix $H$ corresponding to non-zero eigenvalues define
directions of fluctuations of a quantum state and its norm, while eigenvectors
that correspond to zero eigenvalues define directions of insignificant
fluctuations that are due to the ambiguity (\ref{eq:L_implication}) in
definition of the purified state vector.

While information matrix $H$ defines detailed characteristics of precision of
reconstruction of an arbitrary quantum state, the measurement  matrix $B$
(\ref{eq:B}) defines the quality of the protocol as a whole. The condition
number $K$ separates complete and well defined quantum measurement protocols
from incomplete and ill-defined ones.

The approach developed above is based on analysis of completeness, adequacy and
precision of quantum measurements and it could be applied to analysis of
quality and efficiency of arbitrary protocols of quantum measurements.

As a simple example, let us consider a single-qubit protocol proposed in works
\cite{Nunn,Kosut_Walm} (see Fig.~1 in \cite{Nunn} and Fig.~3 in
\cite{Kosut_Walm}). The considered approach is an optimized measurement of
state
$$
c=\frac{1}{\sqrt{2}}\left( \begin{array}{c}
1 \\
1
\end{array} \right).
$$
Measurements are conducted by means of a set of half- and quarter- wave plates
that lead to projections to the following basis states
$$
\psi_1=\frac{1}{\sqrt{2}}\left( \begin{array}{c}
\sin(2h)+i\sin(2(h-q)) \\
\cos(2h)-i\cos(2(h-q))
\end{array} \right).
$$
$$
\psi_2=\frac{1}{\sqrt{2}}\left( \begin{array}{c}
\cos(2h)+i\cos(2(h-q)) \\
-\sin(2h)+i\sin(2(h-q))
\end{array} \right).
$$
where $h$ and $q$ are orientation angles for half-wave and quarter-wave plates
respectively. This measurement defines two strings of instrumental matrix $X$.
The authors of the protocol proposed a set of four measurements of such type (8
rows in total). Measurements are defined by the following orientation angles
for the plates: $h_1=0^\circ$, $q_1=0^\circ$; $h_2=20^\circ$, $q_2=45^\circ$;
$h_3=25^\circ$, $q_3=45^\circ$; $h_4=45^\circ$, $q_4=0^\circ$.

\begin{figure}[b!]
\includegraphics[width=0.95\columnwidth]{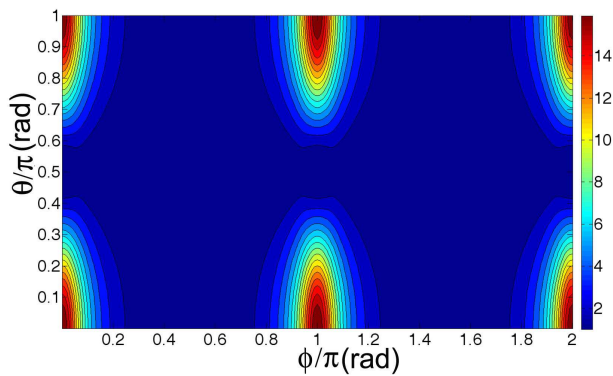}
\caption{(Color online) Analysis of precision losses  for protocol proposed in
\cite{Kosut_Walm,Nunn}.}\label{f:figure20}
\end{figure}

A complete analysis of precision losses (\ref{eq:L}) during reconstruction of
single-qubit pure quantum states using the protocol is given in
Fig.~\ref{f:figure20} (little precision loss states are marked blue, large
precision losses a marked brown). We see that for the state
$$
c=\frac{1}{\sqrt{2}}\left(
\begin{array}{c}
1 \\
1
\end{array} \right)
$$
the protocol is indeed the optimal one because the losses are at minimum
$L_{\text{min}}=1$. However for a number of other states the protocol is rather
not optimal (maximum losses are quite large $L_{\text{max}}\approx 16.84$). For
instance, the protocol considered here has much less precision than those
studied in section~\ref{sec:Discussion} which are based on symmetries of cube
and octahedron and which also have few rows (6 and 8 respectively), but provide
much lower precision losses in narrow range from $L_{\text{min}}=1$ to
$L_{\text{max}}=9/8$.

Note that the condition number $K$ for the protocol is approximately $4.07$
times higher than for protocols based on regular polyhedrons, which also
certifies its lower quality.

\section{Conclusions}\label{sec:Conclude}

In this paper, we proposed a methodology to assess the quality and efficiency
of quantum measurement protocols. It was shown that the proposed approach,
based on the analysis of completeness, adequacy and accuracy, can be
successfully applied to arbitrary quantum states and measurement protocols.
Analysis of the completeness of the protocol allows to answer the question: Is
the set of operators considered sufficient to assess an arbitrary pure or mixed
quantum state? The operational criterion of the protocols' quality follows from
the analysis of completeness. The criterion is based on the condition number of
the measurements matrix $K$. An analysis of the adequacy allows one to answer
the question about the correspondence between the statistical experimental data
and the quantum-mechanical mathematical model. The analysis of the adequacy
guarantees the correctness of the following procedures of statistical quantum
state reconstruction by using the maximum likelihood method. The developed
method of statistical reconstruction is based on on the procedure of quantum
state parameterization using the procedure of purification. Such
parametrization allows us to introduce a universal distribution for Fidelity,
which provides all information about the accuracy of the the quantum state
reconstruction. Various numerical examples and results of physical experiments
have been analyzed for testing theoretical predictions both for pure and mixed
states and demonstrating approach validity.

\section*{Acknowledgments}

This work was supported in part by Russian Foundation of Basic Research
(projects 10-02-00204a, 10-02-00414-a), by Associazione Sviluppo Piemonte and
by Program of Russian Academy of Sciences in fundamental research.


\begin{thebibliography}{99}
\bibitem{instrumental} It holds if one can neglect instrumental uncertainties.
\end{thebibliography}
\end{document}